\documentclass[amsmath,amssymb,prb,twocolumn]{revtex4}
\usepackage[dvips]{graphicx}
\usepackage{bm}

\usepackage{color}

\begin{document}
\title{
Full counting statistics of information content and particle number
}
\author{Yasuhiro Utsumi}
\address{Department of Physics Engineering, Faculty of Engineering, Mie University, Tsu, Mie 514-8507, Japan}

\begin{abstract}
We consider a bipartite quantum conductor and discuss the joint probability distribution of particle number in a subsystem and the self-information associated with the reduced density matrix of the subsystem. 
By extending the multi-contour Keldysh Green function technique, we calculate the R\'enyi entropy of a positive integer order $M$ subjected to the particle number constraint, from which we derive the joint probability distribution. 
For energy-independent transmission, we derive the time dependence of the accessible entanglement entropy, or the conditional entropy. 
We analyze the joint probability distribution for energy-dependent transmission probability at the steady state under the coherent resonant tunneling and the incoherent sequential tunneling conditions. 
We also discuss the probability distribution of the efficiency, which measures the information content transfered by a single electron. 
\end{abstract}

\date{\today}

\maketitle

\newcommand{\mat}[1]{\mbox{\boldmath$#1$}}

\section{Introduction}

Entanglement, a nonlocal correlation existing only between quantum systems~\cite{Einstein,Aharonov}, has been a topic of intensive study since it is essential in quantum information processing~\cite{NC}. 
The prototype of an entangled state is a Bell state, e.g., spatially separated two spin half atoms forming the spin singlet state~\cite{Bohm,Aharonov}. 
About a decade ago, various setups to generate entangled fermions in mesoscopic conductors were proposed~\cite{Beenakker}. 
It has been pointed out that such an entanglement can be created by applying a source-drain bias voltage to a tunnel junction: 
When the applied source-drain bias voltage raises the chemical potential of one lead, e.g., the right lead, an electron moves from the right lead to the left lead and a hole remains in the right lead. 
The electron-hole pair spreading between the left and right leads created in this way can be regarded as a Bell state~\cite{Beenakker1,Beenakker}. 
Several measures for detecting entanglement have been proposed. 
The violation of the Bell inequality~\cite{Bell1964,Clauser1969,Aharonov} can be tested experimentally by measuring correlation functions, i.e., the current noise~\cite{Kawabata,Chtchelkatchev,Samuelsson2004,Faoro}. 
The entanglement witness is another measure and is applied to Kondo systems~\cite{Lee2015}. 

\begin{figure}[ht]
\includegraphics[width=0.5 \columnwidth]{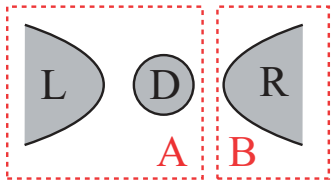}
\caption{
A quantum conductor (a single-level quantum dot) coupled to left and right leads. 
We partition the system into subsystems $A$ and $B$ and trace out the degrees of freedom associated with subsystem $B$. 
Subsystem $A$ consists of the quantum conductor (the dot) and the left lead. 
Subsystem $B$ consists of the right lead. 
}
\label{setup}
\end{figure}

Theoretically, a tractable measure of bipartite entanglement is entanglement entropy~\cite{Beenakker,Klich2009,Klich2}, or R\'enyi entanglement entropy~\cite{FrancisSong1,FrancisSong2,Petrescu2014,Thomas2015}. 
Suppose we partition our system into complementary subsystems $A$ (the left lead and the quantum conductor) and B (the right lead) [Fig.\ref{setup}]. 
Then the partial trace of the density matrix of the total system $\hat{\rho}$ over the subsystem $B$ degrees of freedom, 
\begin{align}
\hat{\rho}_A
=
{\rm Tr}_B
\hat{\rho}
\, , 
\end{align}
defines the reduced density matrix. 
R\'enyi entanglement entropy is given by~\cite{Cover,note1}, 
\begin{align}
S_M
=
{\rm Tr}_A
\left[
{\hat{\rho}_A}^M
\right]
\, , 
\label{renyif}
\end{align}
where ${\rm Tr}_A$ means the partial trace over the subsystem $A$ degrees of freedom. 
The operator of the entanglement spectrum~\cite{Li,Ryu}, 
\begin{align}
\hat{I}_A=-\ln \hat{\rho}_A \, , 
\label{eh}
\end{align}
(we choose base $e$) may be interpreted as the operator of the self-information associated with an outcome described by the reduced density matrix. 
The first derivative of the R\'enyi entropy in terms of $M$ produces the average, i.e., the entanglement entropy, 
\begin{align}
\langle \hat{I}_A \rangle
=
\left.
-\partial_M
S_M
\right|_{M=1}
=
{\rm Tr}_A
\left[
\hat{\rho}_A \hat{I}_A
\right]
\, . 
\label{entent}
\end{align}
In the present paper, we refer to (\ref{entent}) as `full entanglement entropy'~\cite{Klich2}. 
The R\'enyi entanglement entropy of a positive integer order $M$ is, in principle, measurable, by preparing $M$ copies of the total system~\cite{Abanin,Daley}. 
In a bosonic optical lattice, the R\'enyi entropy of order $M=2$, i.e., the purity~\cite{NC}, has been measured experimentally~\cite{Islam}. 
Recent studies have revealed that entanglement entropy and entanglement spectrum are useful for characterizing quantum many-body states~\cite{Amico} and topological states of matter~\cite{Li,Ryu}. 

Several early works~\cite{Wiseman2003,Dowling2006,Beenakker,Klich2} pointed out that the full entanglement entropy defined in Eq.~(\ref{entent}) does not rule out superpositions of different particle number eigenstates, which cannot be created and measured locally. 
Therefore, such superpositions are inaccessible as quantum resources~\cite{Wiseman2003,Beenakker}. 
To take this `local particle number superselection rule'~\cite{Wiseman2003,Dowling2006} into account, we consider the reduced density matrix after measuring the particle number and obtaining the measurement result $N_A$, 
\begin{align}
\hat{\rho}_{A, N_A}
=
\frac{ \hat{\Pi}_{N_A} \hat{\rho}_A \hat{\Pi}_{N_A} }{P(N_A)} 
\, , 
\label{reddenpro}
\end{align}
where $\hat{\Pi}_{N_A}$ is a projection operator onto sectors with electron number ${N_A}$ in the subsystem $A$. 
\begin{align}
P({N_A})
=
{\rm Tr}_A 
\left[ 
\hat{\Pi}_{N_A} \hat{\rho}_A 
\right] 
\, , 
\label{pro}
\end{align}
is the probability of finding $N_A$ particles. 
`Accessible entanglement entropy'~\cite{Wiseman2003,Beenakker,Klich2} is the weighted sum of the entanglement entropy associated with the density matrix (\ref{reddenpro}), 
\begin{align}
\langle J \rangle
=
-
\sum_{N_A}
P({N_A})
{\rm Tr}_A
\left[ 
\hat{\rho}_{A,{N_A}} \ln \hat{\rho}_{A,{N_A}}
\right] \, . 
\label{acceentent} 
\end{align}
We observe that accessible entanglement entropy (\ref{acceentent}) can be regarded as a conditional entropy, which quantifies the average uncertainty associated with the quantum state $\hat{\rho}_A$ after the number of particles $N_A$ is known~\cite{Cover,Klich2}. 
In addition, in our previous work we considered the probability distribution of self-information~\cite{YU2015}. 
Therefore, the above observation motivated us to consider the interplay between the fluctuations of self-information and those of the particle number. 

In the present paper, we consider the R\'enyi entanglement entropy of order $M$ subjected to the particle number constraint, 
\begin{align}
S_M({N_A})
=
{\rm Tr}_A
\left[
(\hat{\Pi}_{N_A} \hat{\rho}_A \hat{\Pi}_{N_A})^M
\right]
\, . 
\label{renyi}
\end{align}
We will relate this to the information generating function~\cite{Golomb,Guiasu}, which is the Fourier transform of the joint probability distribution of self-information and particle number~(\ref{jpdfin}). 
We will extend the multi-contour Keldysh Green function technique~\cite{Nazarov,Ansari1,YU2015} to account for the particle number constraint. 
The advantage of this approach is that the discrete Fourier transform of the multi-contour Keldysh Green function~\cite{YU2015} is reduced to the modified Keldysh Green function~\cite{UGS,BUGS,UEUA,Novotny,Urban,SU2008,US2009,Esposito2009} used in the theory of full counting statistics~\cite{Levitov}. 
Therefore, it is possible to utilize standard Keldysh field theory techniques~\cite{CSHY1985,YU2003,Kamenevbook}.

As an example, we will apply our framework to a simple model, the spinless resonant level model. 
We will present the time dependence of the accessible entanglement entropy and the joint probability distribution of self-information and particle number. 
We will point out that for energy independent transmission, there is a perfect linear correlation between the self-information and the particle number. 
In this case, one can deduce the entanglement entropy by counting the number of transmitted electrons. 

We will further consider an analogy between the information entropy and the thermodynamic entropy. 
Recently, the probability distribution of the efficiency or the coefficient of performance (COP) defined as the fluctuating output work divided by the fluctuating input heat have been discussed~\cite{Verley1,Verley2,Polettini,Jiang,Okada2017,Proesmans2016} in the context of the stochastic thermodynamics~\cite{Seifert}. 
Motivated by these studies, we consider the probability distribution of the COP (\ref{dimlescop}), which measures the information content carried by a single electron. 
It is analogous to the COP of the Peltier effect, which is the fluctuating output heart current divided by the fluctuating input charge current~\cite{Okada2017}. 
We demonstrate the trade-off between the amount of information content carried by a single particle and its uncertainty.

The structure of the paper is the following. 
In Sec.~\ref{jpdf}, we will introduce the joint probability distribution of self-information and particle number and the probability distribution of conditional self-information. 
In Sec.~\ref{RenPar}, will present the replica method for calculating the R\'enyi entropy of integer order subjected to the particle number constraint.
In Sec.~\ref{mkgf}, we will introduce the spinless resonant level model as a simple example. 
Then we summarize the multi-contour Keldysh Green function modified with the particle number counting field. 
In Sec.~\ref{srlm}, we present an explicit form of the R\'enyi entropy subjected to the particle number constraint. 
We analyze the time dependence of the accessible entanglement entropy and the probability distribution of conditional self-information. 
In Sec.~\ref{edtp}, we analyze the joint probability distribution of self-information and particle number at the steady state. 
Then we discuss the analogy between the thermoelectric effect, the Peltier effect, and our information transmission setup in Sec.~\ref{peltier}. 
Section~\ref{summary} summarizes our results.

\section{Full counting statistics}
\label{jpdf}

\subsection{Joint probability distribution}

After the projection measurement of the number of particles in subsystem $A$, the self-information operator (\ref{eh}) may be modified as, 
\begin{align}
\hat{I}_A^\prime 
= 
-\ln \hat{\rho}_{A}^\prime 
\, , 
\label{ehss}
\end{align}
where 
\begin{align}
\hat{\rho}_{A}^\prime = \sum_{N_A} \hat{\Pi}_{N_A} \hat{\rho}_A \hat{\Pi}_{N_A}
\, . 
\end{align}
The prime indicates that the operator is written with the density matrix after the projection measurement. 
The joint probability distribution to find $I_A'$ and ${N_A}$ may be written as,
\begin{align}
P(I_A',{N_A})
=
{\rm Tr}_A
\left[ 
\hat{\Pi}_{N_A} \hat{\rho}_{A} \hat{\Pi}_{N_A}
\delta(I_A'-\hat{I}_A^\prime)
\right]
\, . 
\label{jpdfin}
\end{align}
The information generating function~\cite{Golomb}, i.e., the characteristic function of the probability distribution of self-information subjected to the particle number constraint, is 
\begin{align}
S_{1- i \xi}({N_A})
=&
\int d I_A^\prime e^{i \xi I_A^\prime}
P(I_A^\prime,{N_A})
\nonumber \\
=&
{\rm Tr}_A
\left[
\left(
\hat{\Pi}_{{N_A}} \hat{\rho}_A \hat{\Pi}_{{N_A}}
\right)^{1-i \xi}
\right]
\, . 
\label{igf}
\end{align}
The information generating function is obtained from the R\'enyi entropy (\ref{renyi}) by extending $M$ to $1-i \xi$. 
The R\'enyi entropy (\ref{renyi}) is related to the probability (\ref{pro}) as, 
\begin{align}
P(N_A)
=
S_{1}(N_A)
\, . 
\label{ren2pro}
\end{align}
We perform the Fourier transform in $N_A$ and introduce the information generating function, which we refer to as the modified R\'enyi entropy, 
\begin{align}
S_M(\chi)
=
\sum_{N_A}
S_M(N_A)
\, 
e^{i \chi N_A}
=
{\rm Tr}_A
\left(
e^{i \chi \hat{N}_A}
\hat{\rho}_A^{\prime \; \, M}
\right)
\, , 
\label{igfbi}
\end{align}
where $\hat{N}_A$ is the number operator of electrons in subsystem $A$. 

A joint moment is calculated by the derivative of the information generating function with respect to the counting fields $\chi$ and $\xi$ as, 
\begin{align}
\langle 
{\hat{I}_A}^{\prime \;\, \ell}
{\hat{N}_A}^m
\rangle
=&
{\rm Tr}_A
\left(
{\hat{\rho}_A^\prime}
{\hat{I}_A}^{\prime \;\, \ell}
{\hat{N}_A}^m
\right)
\nonumber \\
=&
\left.
{\partial_{i \xi}}^\ell
{\partial_{i \chi}}^m
S_{1-i \xi}(\chi)
\right|_{\chi=\xi=0}
\, . 
\end{align}
Since the self-information operator and the particle number operator commute $[\hat{I}_A^\prime, \hat{N}_A]=0$, it is a classical correlation function. 
Similarly, a joint cumulant is, 
\begin{align}
\langle \! \langle 
{{I}_A'}^\ell
{{N}_A}^m
\rangle \! \rangle
=
\left.
{\partial_{i \xi}}^\ell
{\partial_{i \chi}}^m
\ln 
S_{1-i \xi}(\chi)
\right|_{\chi=\xi=0}
\, . 
\label{joicum}
\end{align}
The explicit form of the first cumulant, the average of the self-information operator (\ref{ehss}), is 
\begin{align}
\langle \! \langle 
{{I}_A'}
\rangle \! \rangle
=
\langle 
\hat{I}_A^\prime
\rangle
=
-
{\rm Tr}_A
\left(
\hat{\rho}_A^\prime
\ln 
\hat{\rho}_A^\prime
\right)
\, . 
\label{je}
\end{align}
This can be regarded as the joint entropy, which measures the average uncertainty associated with the quantum state $\hat{\rho}_A$ and the particle number $N_A$~\cite{NC,Cover}. 
Expressions of second joint cumulants, the variance and the covariance, are 
\begin{subequations}
\begin{align}
\langle \! \langle 
{{I}_A^\prime}^2
\rangle \! \rangle
=&
\langle 
{\delta \hat{I}_A^\prime}^2
\rangle
\, ,
\\
\langle \! \langle 
{{N}_A}^2
\rangle \! \rangle
=&
\langle 
{\delta \hat{N}_A}^2
\rangle
\, ,
\\
\langle \! \langle 
{{I}_A'}
{{N}_A}
\rangle \! \rangle
=&
\langle 
\delta { \hat{I}_A' }
\delta {\hat{N}_A }
\rangle
\, , 
\end{align}
\end{subequations}
where
$\delta \hat{I}_A^\prime=\hat{I}_A^\prime-\langle \hat{I}_A^\prime \rangle$ 
and 
$\delta \hat{N}_A=\hat{N}_A-\langle \hat{N}_A \rangle$. 
The correlation coefficient $r$, which measures how two fluctuating quantities ${I}_A^\prime$ and ${N}_A$ are linearly correlated, is defined as~\cite{Hogg,Okada2017} 
\begin{align}
-1
\leq 
r=
\frac{
\langle \! \langle 
{{N}_A}
{{I}_A^\prime}
\rangle \! \rangle
}{
\sqrt{
\langle \! \langle 
{{I}_A^\prime}^2
\rangle \! \rangle
\langle \! \langle 
{{N}_A}^2
\rangle \! \rangle
}
}
\leq 1
\, . 
\label{corcoe}
\end{align}
The correlation coefficient takes the maximum (minimum) value $1$ ($-1$) when two quantities are linearly dependent $I_A'= \alpha N_A + {\rm const}.$ 
with positive (negative) slope $\alpha$~\cite{note2}. 
Therefore, when $r=\pm1$, there exists a one-to-one correspondence between the self-information and the particle number. 
In such a case, one can deduce the entanglement entropy by measuring the particle number.

\subsection{Probability distribution of conditional self-information}
\label{conselinf}

Once we measure the particle number of the subsystem $A$ and obtained the result $N_A$, the uncertainty is reduced by $-\ln P(N_A)$. 
The conditional self-information $J$ is the self-information under the condition that $N_A$ is known. 
Its probability distribution may be defined by utilizing the joint probability distribution function (\ref{jpdfin}) as, 
\begin{align}
P(J)
=
\sum_{{N_A}}
\int d I_A'
P(I_A',{N_A})
\delta \left( J-I_A'-\ln P({N_A}) \right)
\, , 
\label{pdfci}
\end{align}
Then the information generating function~\cite{Golomb,Guiasu} becomes, 
\begin{align}
R_{1-i \xi}
=
\int d J e^{i \xi J}
P(J)
=
\sum_{{N_A}}
S_{1-i \xi}({N_A})
S_{1}({N_A})^{i \xi}
\, . 
\label{rigf}
\end{align}
The $n$th cumulant is calculated as, 
\begin{align}
\langle \! \langle J^n \rangle \! \rangle
=
\left.
\partial_{i \xi}^n
\ln R_{1-i \xi}
\right|_{\xi=0}
\, . 
\label{jcum}
\end{align}
Here we utilized the normalization condition, see Eq.~(\ref{ncr}). 
One can check that the first derivative reproduces the accessible entanglement entropy (\ref{acceentent}), which is also rewritten in the following form~\cite{Klich2}; 
\begin{align}
\langle \! \langle J \rangle \! \rangle
=
\langle \! \langle I_A' \rangle \! \rangle
-
H({N_A}) 
\, . 
\label{ententsupsel1} 
\end{align}
This is the chain rule~\cite{NC,Cover} connecting the conditional entropy to the joint entropy (\ref{je}) and the Shannon entropy associated with the probability distribution of the number of particles (\ref{pro});  
\begin{align}
H({N_A})
=
-
\sum_{N_A}
P({N_A})
\ln P({N_A})
\, . 
\label{Shannonentropy}
\end{align}

We remark that the joint entropy (\ref{je}) is equivalent to the full entanglement entropy (\ref{entent}) for certain situations. 
For nonsuperconducting leads, the conservation of the {\it total} particle number ensures that the reduced density matrix at time $\tau$, $\hat{\rho}_A(\tau)$, and the local particle number operator $\hat{N}_A$ commute (Appendix~\ref{sec:prolocssr}), 
\begin{align}
[ \hat{\rho}_A(\tau), \hat{N}_A ] =0 \, . 
\label{pnsc}
\end{align}
Therefore, the reduced system is always a statistical mixture of states with different particle numbers. 
Since Eq.~(\ref{pnsc}) implies $\hat{\rho}_A^\prime = \hat{\rho}_A$, the joint entropy (\ref{je}) is equal to the full entanglement entropy (\ref{entent}); 
$\langle \hat{I}_A^\prime \rangle = \langle \hat{I}_A \rangle$. 
By combining it with Eq.~(\ref{ententsupsel1}), one can relate the accessible entanglement entropy to the full-entanglement entropy~\cite{Klich2} as, 
\begin{align}
\langle \! \langle J \rangle \! \rangle
=
\langle I_A \rangle
-
H({N_A}) 
\, . 
\label{ententssr2} 
\end{align}

\subsection{Universal relations}
\label{sec:normalization}

The modified R\'enyi entropy (\ref{igfbi}) and the information generating function (\ref{rigf}) satisfy the normalization condition, 
\begin{align}
S_1(\chi=0)
=&
\sum_{N_A=-\infty}^\infty
\int d I_A^\prime
P(I_A^\prime,N_A) 
=1
\, , 
\\
R_1
=&
\int dJ P(J)
=
\sum_{N_A} S_1({N_A})
=1
\, . 
\label{ncr}
\end{align}

There exist universal relations formally similar to the Jarzynski equality~\cite{Jarzynski1997,Campisi2011}. 
The R\'enyi entropy~(\ref{renyif}) satisfies a universal relation [Eq.~(8) in Ref.~\onlinecite{YU2015}], 
\begin{align}
S_0=
\langle e^{I_A} \rangle
=
\int d I_A P(I_A) e^{I_A}
=
{\rm rank} \left( \hat{\rho}_A \right)
\, , 
\label{jarequ2015}
\end{align}
which is the size of available states in the Fock space. 
There exist universal relations associated with the information generating functions (\ref{igf}) and (\ref{rigf}). 
The information generating function (\ref{igf}) satisfies, 
\begin{align}
S_0(N_A)=
\int d I_A^\prime e^{I_A^\prime}
P(I_A^\prime,{N_A})
=
{\rm rank} \left( \hat{\Pi}_{N_A} \hat{\rho}_A \hat{\Pi}_{N_A} \right)
\, ,
\label{jarequ}
\end{align}
which is the size of available states in the Fock subspace containing $N_A$ particles. 
The information generating function (\ref{rigf}) satisfies, 
\begin{align}
R_0
=&
\left \langle e^J \right \rangle
=
\int dJ e^J P(J)
=
\sum_{N_A}
S_0(N_A)
P({N_A})
\, . 
\label{jaeqr}
\end{align}
If $P(J) \geq 0$, by applying Jensen's inequality to Eq.~(\ref{jaeqr}), 
we obtain the `second law of thermodynamics'~\cite{Jarzynski1997,Campisi2011} for accessible entanglement entropy, 
\begin{align}
\langle J \rangle
\leq 
\ln \sum_{N_A} S_0(N_A) P({N_A}) \, . 
\end{align}

\section{R\'enyi entropy subjected to a particle number constraint}
\label{RenPar}

\subsection{Replica method}

\begin{figure}[hb]
\includegraphics[width=0.8 \columnwidth]{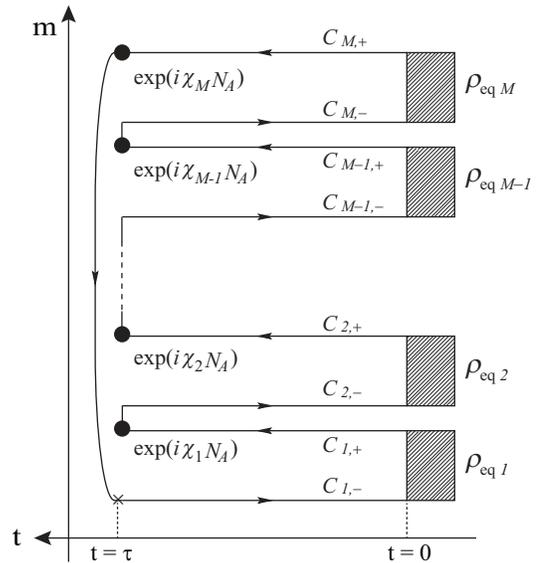}
\caption{
Multi-contour $C$ consisting of $M$ replicated standard Keldysh contours. 
The cross at $t=\tau$ on the lower branch of the first replica $C_{1 , -}$ represents the starting point. 
Shaded boxes are $M$ replicas of the initial equilibrium density matrix $\hat{\rho}_{{\rm eq} \, m}$ ($m=1,\cdots,M$). 
Solid circles on $t=\tau_{m+}$ represent the operator $\exp(i \chi_m \hat{N}_A)$. 
}
\label{rkc}
\end{figure}

In this section, we introduce the multi-contour $C$ and express the R\'enyi entropy of a positive integer order $M$ subjected to the particle number constraint~(\ref{renyi}) by exploiting it. 
The particle-number projection operator is, 
\begin{align}
\hat{\Pi}_{N_A}
=
\int_{-\pi}^\pi \frac{d \chi}{2 \pi}
e^{i (\hat{N}_A-{N_A}) \chi}
\, . 
\end{align}
By utilizing the property, ${\hat{\Pi}_{N_A}}^2=\hat{\Pi}_{N_A}$, 
the modified R\'enyi entropy~(\ref{igfbi}) is expressed as, 
\begin{align}
S_M(\chi)
=&
\int_{-\pi}^\pi
\frac{d \chi_M}{2 \pi}
\cdots 
\frac{d \chi_1}{2 \pi}
2 \pi \delta(\chi - \bar{\chi})
\, 
S_M(\{ \chi_m \})
\, , 
\label{renss}
\end{align}
where 
$\bar{\chi}=\sum_{m=1}^M \chi_m$ 
and, 
\begin{align}
S_M(\{ \chi_m \})
=&
S_M(\chi_1,\cdots,\chi_M)
\nonumber \\
=&
{\rm Tr}_A
\left[
e^{i \chi_M \hat{N}_A}
\hat{\rho}_A
\cdots 
e^{i \chi_1 \hat{N}_A}
\hat{\rho}_A
\right]
\, . 
\label{rensscou}
\end{align}
Averages over phases between adjacent reduced density matrices imply the lack of a particular phase reference between subsystems $A$ and $B$~\cite{Dowling2006}.

The full density matrix at time $\tau$ is prepared in the following manner. 
Following the standard procedure of non-equilibrium quantum transport problems (see, e.g., Ref.~\onlinecite{Meir}), we assume that initially subsystems $A$ and $B$ are decoupled and that each subsystem is in equilibrium. 
Then the initial full density matrix is, 
\begin{align}
\hat{\rho}_{{\rm eq}}
= 
\hat{\rho}_{A {\rm eq}} \hat{\rho}_{B {\rm eq}}
\, . 
\label{rhoini}
\end{align}
The full density matrix evolves during time $\tau$ as, 
\begin{align}
\hat{\rho}(\tau)
= 
\hat{U}(\tau)  \hat{\rho}_{{\rm eq}} \hat{U}(\tau)^\dagger
\, ,
\;\;\;\;
\hat{U}(\tau) = e^{-i \hat{H} \tau}
\, , 
\label{rhofull}
\end{align}
where $\hat{H}=\hat{H}_0 + \hat{V}$. 
The nonperturbative Hamiltonian is 
$\hat{H}_0 = \hat{H}_A + \hat{H}_{B}$, 
where $\hat{H}_A$ and $\hat{H}_{B}$ are Hamiltonians of the subsystem $A$ and the subsystem $B$, respectively. 
The perturbation Hamiltonian $\hat{V}$ describes the coupling between the two subsystems. 

For a positive integer $M$, the modified R\'enyi entropy~(\ref{rensscou}) is calculated~\cite{YU2015} by exploiting the replica method~\cite{Sherrington1975} and the multi-contour Keldysh Green function technique~\cite{Nazarov,Ansari1}. 
We introduce $M$ replicas of the subsystem $B$ and introduce operators associated with $m$th replica ($m=1,\cdots,M$) as, 
\begin{align}
\hat{H}_{B} & \to \hat{H}_{B,m} \, , \nonumber \\
\hat{V} & \to \hat{V}_m \, , \nonumber  \\
\hat{\rho}_{B {\rm eq}} & \to \hat{\rho}_{B {\rm eq},m} 
\, . 
\label{repope}
\end{align}
The $m$th full density matrix is 
$\hat{\rho}_{m} = \hat{U}_m \hat{\rho}_{\rm eq,m} \hat{U}_m^\dagger$, 
where 
$\hat{\rho}_{\rm eq,m} = \hat{\rho}_{A {\rm eq}} \hat{\rho}_{B {\rm eq},m}$. 
The $m$th replicated time evolution operator is, 
$\hat{U}_m = e^{-i \hat{H}_m \tau}$, 
where 
$\hat{H}_m = \hat{H}_{0,m} + \hat{V}_{m}$ 
and 
$\hat{H}_{0,m} = \hat{H}_{A} + \hat{H}_{B,m}$. 
The $m$th replicated reduced density matrix is obtained from the $m$th full density matrix $\hat{\rho}_{m}$ by tracing out the degrees of freedom of the $m$th replicated subsystem $B$; 
$\hat{\rho}_{A,m} = {\rm Tr}_{B,m}[\hat{\rho}_m]$. 
Then the modified R\'enyi entropy~(\ref{rensscou}) is, 
\begin{align}
S_M(\{ \chi_m \})
=&
{\rm Tr}_A
\left[
e^{i \chi_M \hat{N}_A}
\hat{\rho}_{A,M}
\cdots 
e^{i \chi_1 \hat{N}_A}
\hat{\rho}_{A,1}
\right]
\nonumber \\
=&
{\rm Tr}
\left[
e^{i \chi_M \hat{N}_A}
\hat{\rho}_{M}
\cdots 
e^{i \chi_1 \hat{N}_A}
\hat{\rho}_{1}
\right]
\, . 
\label{renssrep}
\end{align}
Here the trace in the second line should be performed over the total system, the subsystem $A$ and $M$ replicas of subsystem $B$; 
${\rm Tr}={\rm Tr}_A {\rm Tr}_{B,M} \cdots {\rm Tr}_{B,1}$. 
The modified R\'enyi entropy is rewritten as, 
\begin{align}
S_M( \{ \chi_m \} )
=&
{\rm Tr}
\left[
e^{i \chi_M \hat{N}_{A} }
\hat{U}_M
\hat{\rho}_{{\rm eq} \, M}
\hat{U}_M^\dagger
\right.
\nonumber \\
& 
\left.
\times
\cdots
e^{i \chi_1 \hat{N}_{A} }
\hat{U}_{1}
\hat{\rho}_{{\rm eq} \, 1}
\hat{U}_{1}^\dagger
\right]
\, , 
\end{align}
which is visualized in Fig.~\ref{rkc}. 
Following Ref.~\onlinecite{Nazarov}, we introduce the multi-contour $C$, which is a sequence of $M$ standard Keldysh contours. 
We set a starting point at $t=\tau$ on the lower branch of the first Keldysh contour $C_{1 , -}$ (cross in Fig.~\ref{rkc}). 
The contour goes to $\hat{\rho}_{{\rm eq} \, 1}$ at $t=0$ along $C_{1 , -}$ and returns to $t=\tau$ along $C_{1 , +}$. 
Then it connects to $t=\tau$ on the lower branch of the second Keldysh contour $C_{2 , -}$. 
The contour goes repeatedly until it reaches $t=\tau$ on $C_{M , +}$. 
Then it goes back to the starting point $t=\tau$ on $C_{1 , -}$. 

In the interaction picture, the time evolution operator and its Hermitian conjugate are expanded as 
$
\hat{U}_{m \, I}
=
e^{i \hat{H}_{0 \, m} \tau} {U_m}
=
\hat{T} 
\exp
\left(
-i \int_0^\tau dt 
\hat{V}_{m}(t)_I
\right)
$
and 
$
\hat{U}_{m \, I}^\dagger
=
\hat{ \tilde{T} }
\exp
\left(
i \int_0^\tau dt 
\hat{V}_{m}(t)_I
\right)
\, , 
$
where $\hat{T}$ ($\hat{\tilde{T}}$) is the (anti-) time-ordering operator. 
The perturbation Hamiltonian in the interaction picture is 
$
\hat{V}_{m}(t)_I
=
e^{i \hat{H}_{0 \, m} t}
\hat{V}_{m}
e^{-i \hat{H}_{0 \, m} t}
$. 
By exploiting the multi-contour $C$ and the contour ordering operator $\hat{T}_C$, 
the modified R\'enyi entropy is expressed as,
\begin{align}
\frac{
S_M(\{ \chi_m \})
}{s_M(\bar{\chi})}
=
\left \langle 
\hat{T}_C
e^{
-i \int_C dt 
\hat{V}(t)_I
+
i
\sum_{j=1}^M
\chi_j 
\hat{N}_{A}(\tau_{j +})_I
}
\right \rangle_M 
\, , 
\label{kpf}
\end{align}
which is the `Keldysh partition function'. 
The angle brackets stand for the expectation value, 
\begin{align}
\langle \hat{\mathcal O} \rangle_M
=
{\rm Tr}
\left[
\hat{\mathcal O} 
\hat{\rho}_{{\rm eq} \, M}
\cdots
\hat{\rho}_{{\rm eq} \, 1}
\right]
/s_M(\bar{\chi})
\, , 
\label{nev}
\end{align}
normalized by the modified R\'enyi entropy of decoupled subsystems, 
\begin{align}
s_M(\bar{\chi})
=&
{\rm Tr}_A
\left[
e^{i \bar{\chi} \hat{N}_A}
{\hat{\rho}_{A, {\rm eq}}}^M
\right]
\label{rendec}
\, . 
\end{align}
In Eq.~(\ref{kpf}), the integral over $t$ is performed along the multi-contour $C$. 
We wrote the time $\tau$ defined on $C_{m , \pm}$ as $\tau_{m \pm}$. 
The contour-ordering operator $\hat{T}_C$ also acts on $\hat{\rho}_{{\rm eq} \, m} $ residing on $t=0_ {m \pm}$. 

\subsection{Global constraint}

In order to perform the multiple integral in Eq.~(\ref{renss}), we transform integration variables from 
$ d\chi_1 \cdots d \chi_{M-1} d \chi_M$
to 
$d \delta \chi_1 \cdots d \delta \chi_{M-1} d \bar{\chi}$
where
$
\delta \chi_m
=
\chi_m
-
\bar{\chi}/M
$ 
$(m=1,\cdots,M-1)$. 
The Jacobian of this transformation is 1. 
Because of the discreteness of electrons, the integrand is expected to be $2 \pi$ periodic in $\chi$ in Eq.~(\ref{renss}). 
Then the modified R\'enyi entropy (\ref{renss}) can be written as, 
\begin{widetext}
\begin{align}
S_M(\chi)
=
\int_{-\pi}^\pi
\frac{d \delta \chi_1}{2 \pi}
\cdots 
\frac{d \delta \chi_{M-1}}{2 \pi}
S_M \!
\left(
\frac{\chi}{M} + \delta \chi_1,
\cdots
\frac{\chi}{M} + \delta \chi_{M-1},
\frac{\chi}{M} - \sum_{j=1}^{M-1} \delta \chi_j
\right)
\, . 
\end{align}
\end{widetext}
The local constraint, the multiple integral over $\delta \chi_j$ ($j=1,\cdots,M-1$), removes the coherence between Fock subspaces with different particle numbers. 
This can be done easily for nonsuperconducting leads, since by exploiting Eq.~(\ref{pnsc}), one immediately sees that the modified R\'enyi entropy~(\ref{rensscou}) depends only on $\bar{\chi}$ and is independent of $\delta \chi_j$. 
Thus we only take the global constraint $\bar{\chi}=\chi$ into account, 
\begin{align}
S_M
\left( \chi \right)
=&
S_M
\left( \{ \chi/M \} \right)
\nonumber \\
=&
{\rm Tr}
\left[
e^{i \chi \hat{N}_A/M}
\hat{\rho}_M
\cdots
e^{i \chi \hat{N}_A/M}
\hat{\rho}_1
\right]
\, . 
\label{renyiglobal}
\end{align}

\section{Modified multi-contour Keldysh Green function}
\label{mkgf}

\subsection{Spinless resonant level model}

We will consider a simple Hamiltonian, the spinless resonant level model, 
and introduce the multi-contour Keldysh Green function. 
Figure \ref{setup} is a schematic picture of our setup. 
We bipartite the system and regard the dot and the left lead as subsystem $A$ and the right lead as subsystem $B$. 
The Hamiltonians of the two subsystems are, 
$\hat{H}_A = \hat{H}_L + \hat{H}_D$ and $\hat{H}_B = \hat{H}_R$. 
The quantum dot is represented by a localized level with the energy $\epsilon_D$, 
$\hat{H}_D = \epsilon_D \hat{d}^\dagger \hat{d}$, 
where $\hat{d}$ is an electron annihilation operator. 
The left ($r=L$) and right ($r=R$) leads are described by the free electron gas; 
$
\hat{H}_r = 
\sum_{k} \epsilon_{rk} \hat{a}_{r k}^\dagger \hat{a}_{r k} 
$, 
where $\hat{a}_{r k}$ annihilates an electron with the wave number $k$ in the lead $r$. 
The coupling between the two subsystems is described by the tunneling Hamiltonian; 
$
\hat{V} = 
\sum_{r=L,R} 
\sum_{k} 
J_r \hat{d}^\dagger \hat{a}_{r k} + {\rm H.c.} 
$. 
The initial density matrices of subsystems $A$ and $B$ are, 
$
\hat{\rho}_{A, {\rm eq}} = \hat{\rho}_{L, {\rm eq}} \hat{\rho}_{D, {\rm eq}}
$
and 
$
\hat{\rho}_{B, {\rm eq}} = \hat{\rho}_{R, {\rm eq}}
$, 
where the density matrix of each region is, 
\begin{align}
\hat{\rho}_{s, {\rm eq}}
=
\frac{
e^{-\beta (\hat{H}_s - \mu_s \hat{N}_s)}
}{
{\rm Tr}_s
\left[
e^{-\beta (\hat{H}_s - \mu_s \hat{N}_s)}
\right]
}
\, , 
\;\;\;\;
(s=L,D,R)
\, . 
\end{align}
We set the chemical potentials as $\mu_A=\mu_L=\mu_D$ and $\mu_B=\mu_R$. 
The number operators of particles in subsystems $A$ and $B$ are, 
$\hat{N}_A = \hat{N}_L + \hat{N}_D$ and $\hat{N}_B = \hat{N}_R$, respectively. 
Here the number operators of particles in the lead $r$ and the dot are, 
$
\hat{N}_r = \sum_{k} \hat{a}_{r k}^\dagger \hat{a}_{r k}
$, 
and 
$\hat{N}_D = \hat{d}^\dagger \hat{d}
$, 
respectively. 
Replicated operators (\ref{repope}) are introduced by replacing 
$\hat{a}_{Rk}$
with 
$\hat{a}_{Rk m}$. 


The modified R\'enyi entropy of decoupled subsystems $\hat{V}=0$ (\ref{rendec}) is 
$s_M(\bar{\chi}) = s_{D \, M}(\bar{\chi}) \, s_{L \, M}(\bar{\chi})$, 
where the dot part is 
\begin{align}
s_{D \, M}(\bar{\chi})
=&
{\rm Tr}_D
\left[
e^{i \bar{\chi} \hat{N}_D}
{ \hat{\rho}_{D, {\rm eq}} }^M
\right]
\nonumber \\
=&
f_D^-(\epsilon_D)^M
+
f_D^+(\epsilon_D)^M
e^{i \bar{\chi}}
\, . 
\label{rendotchi} 
\end{align}
The electron and hole distribution functions are, 
\begin{align}
f_{L}^+ (\epsilon)
=
\frac{1}
{1+e^{\beta (\epsilon-\mu_L)} }
\, ,
\;\;\;\;
f_{L}^- (\epsilon)
=
1-f_{L}^+ (\epsilon)
\, . 
\end{align}
Similarly, the left lead part is, 
$
s_{L \, M}(\bar{\chi})
=
\prod_k
\left[
f_L^-(\epsilon_{Lk})^M
+
f_L^+(\epsilon_{Lk})^M
e^{i \bar{\chi}}
\right]
$.
The order between taking the zero-temperature limit and performing an analytic continuation matters, since $0^0$ is indeterminate~\cite{YU2015} (Appendix \ref{sec:renentdec}). 
In the present paper, we first take the zero-temperature limit for a positive integer $M$, and then extend $M$ to a complex number. 
Then the modified R\'enyi entropy of decoupled subsystems is, 
\begin{align}
s_{M}(\bar{\chi})
=
e^{i \bar{\chi} N_{A,0}}
\, , 
\label{modrendec}
\end{align}
where $N_{A,0}$ is the number of electrons in the subsystem $A$ at 0K; 
$ N_{A,0} = \theta(\mu_D-\epsilon_D) + \sum_{k} \theta(\mu_L-\epsilon_{Lk}) $. 

\subsection{Modified multi-contour Keldysh Green function}

In order to perform a perturbation expansion of the Keldysh partition function (\ref{kpf}), we introduce the subsystem $A$ multi-contour Keldysh Green function modified with the particle number counting field $\chi_m$. 
In the following, we consider the left lead Green function. 
The dot Green function is introduced in the same manner. 
The modified multi-contour Keldysh Green function of $\hat{a}_{Lk}^\dagger$ on $C_{m' , s'}$ and $\hat{a}_{Lk}$ on $C_{m , s}$ is defined as, 
\begin{widetext}
\begin{align}
g_{L k}^{  \{\chi_j \} }(t_{ms},t'_{m's'})
=
g_{L k}^{ \{\chi_j \}, ms,m's'}(t,t')
=
-i 
\left \langle 
T_C
e^{i \sum_{j=1}^M \chi_j \hat{N}_A(\tau_{j+})_I}
\hat{a}_{L k}(t_{ms})_I
\hat{a}_{L k}^\dagger(t_{m's'}')_I
\right \rangle_M
\, . 
\label{gf}
\end{align}
This is a component of a $2M \times 2M$ Keldysh Green function matrix ${\mathbf g}_{L k}(t,t')$. 
A $2 \times 2$ sub-matrix of a $2M \times 2M$ multi-contour Keldysh Green function matrix connecting branches $C_{m, \, \pm}$ and $C_{m' \, \pm}$ reads (Appendix~\ref{sec:mcgf}), 
\begin{align}
\left[
{\mathbf g}_{L k}^{ \{ \chi_j \} }(t,t')
\right]_{m,m'}
=&
\left[
\begin{array}{cc}
g_{L k}^{\{ \chi_j \} , m+,m'+}(t,t') & g_{L k}^{\{ \chi_j \} , m+,m'-}(t,t') \\
g_{L k}^{\{ \chi_j \} , m-,m'+}(t,t') & g_{L k}^{\{ \chi_j \} , m-,m'-}(t,t')
\end{array}
\right]
=
-i 
e^{-i \epsilon_{Lk} (t-t')}
\nonumber \\
&\times
\left \{
\begin{array}{cc}
e^{i \sum_{j=m'}^{m-1} \delta \chi_j}
\left[
\begin{array}{cc}
f^{\bar{\chi}}_{L,m-m'}(\epsilon_{Lk}) & f^{\bar{\chi}}_{L,m-m'+1}(\epsilon_{Lk}) e^{-i {\bar{\chi}}/M} \\
f^{\bar{\chi}}_{L,m-m'-1}(\epsilon_{Lk}) e^{i {\bar{\chi}}/M} & f^{\bar{\chi}}_{L,m-m'}(\epsilon_{Lk})
\end{array}
\right]
& 
(m > m')
\\
\left[
\begin{array}{cc}
f^{\bar{\chi}}_{L,0}(\epsilon_{Lk}) \theta(t-t') - f^{\bar{\chi}}_{L,M}(\epsilon_{Lk}) \theta(t'-t) 
& 
f^{\bar{\chi}}_{L,1}(\epsilon_{Lk}) e^{-i {\bar{\chi}}/M}
\\
-
f^{\bar{\chi}}_{L,M-1}(\epsilon_{Lk}) e^{i {\bar{\chi}}/M}
& f^{\bar{\chi}}_{L,0}(\epsilon_{Lk}) \theta(t'-t) - f^{\bar{\chi}}_{L,M}(\epsilon_{Lk}) \theta(t-t')
\end{array}
\right]
&
(m'=m)
\\
e^{-i \sum_{j=m}^{m'-1} \delta \chi_j}
\left[
\begin{array}{cc}
-f^{\bar{\chi}}_{L,M+m-m'}(\epsilon_{Lk}) & -f^{\bar{\chi}}_{L,M+m-m'+1}(\epsilon_{Lk}) e^{-i {\bar{\chi}}/M} \\
-f^{\bar{\chi}}_{L,M+m-m'-1}(\epsilon_{Lk}) e^{i {\bar{\chi}}/M} & -f^{\bar{\chi}}_{L,M+m-m'}(\epsilon_{Lk})
\end{array}
\right]
&
(m<m')
\end{array}
\right.
. 
\label{matmcgf}
\end{align}
\end{widetext}
where the modified Fermi distribution function is, 
\begin{align}
f^{\bar{\chi}}_{L, m}(\epsilon_{Lk})
=
\frac{
e^{m [i \bar{\chi}/M- \beta (\epsilon_{Lk}-\mu_L)]}
}
{1+
e^{i \bar{\chi}- M \beta (\epsilon_{Lk}-\mu_L)}
}
\, . 
\end{align}
For $\chi_m=0$ ($m=1,\cdots,M$), Eq.~(\ref{matmcgf}) is reduced to Eq. (B3) in Ref.~\onlinecite{YU2015}.

As we discussed, when the reduced density matrix and the particle number operator of subsystem $A$ commute [see Eq.~(\ref{pnsc})], the modified R\'enyi entropy (\ref{renyiglobal}) is independent of $\delta \chi_m$. 
Therefore, we can set $\delta \chi_m=0$. 
Then the $(m,m')$-component of the modified multi-contour Keldysh Green function depends only on $m-m'$, and we utilize the following discrete Fourier transform~\cite{YU2015} (Appendix~\ref{sec:dft}), 
\begin{subequations}
\begin{align}
{\mathbf g}_{L k}^{\lambda_\ell-\chi/M}(t,t')
=&
\sum_{m-m'=0}^{M-1}
\left[
{\mathbf g}_{L k}^{ \{ \chi/M \} }(t,t')
\right]_{m,m'}
\nonumber \\
\times &
e^{i (\pi - \lambda_\ell) (m-m')}
\label{dft}
\, , 
\\
\lambda_\ell =& \pi \left( 1-\frac{2 \ell+1}{M} \right)
\, . 
\end{align}
\end{subequations}
Then the $2 M \times 2 M$ Green function matrix is reduced to the $2 \times 2$ Green function matrix defined on the standard Keldysh space;  
\begin{widetext}
\begin{align}
{\mathbf g}_{L k}^{\lambda}(t,t')
=&
\left[
\begin{array}{cc}
g_{L k}^{\lambda, ++}(t,t') & g_{L k}^{\lambda, +-}(t,t') \\
g_{L k}^{\lambda, -+}(t,t') & g_{L k}^{\lambda, --}(t,t')
\end{array}
\right]
\nonumber \\
=&
-i e^{-i \epsilon_{Lk}(t-t')}
\left[
\begin{array}{cc}
f_{L,\lambda}^-(\epsilon_{Lk})
\theta (t-t')
-
f_{L,\lambda}^+(\epsilon_{Lk})
\theta (t'-t)
&
f_{L,\lambda}^+(\epsilon_{Lk}) e^{i \lambda} \\
-
f_{L,\lambda}^-(\epsilon_{Lk}) e^{-i \lambda}
&
f_{L,\lambda}^-(\epsilon_{Lk})
\theta (t'-t)
-
f_{L,\lambda}^+(\epsilon_{Lk})
\theta (t-t')
\end{array}
\right]
\, . 
\label{mmkgf}
\end{align}
\end{widetext}
The modified electron and hole distribution functions are 
\begin{align}
f_{L,\lambda}^+(\epsilon)
=
\frac{1}
{1+e^{\beta (\epsilon-\mu_L) + i \lambda} }
\, ,
\;\;\;\;
f_{L,\lambda}^-(\epsilon)
=
1
-
f_{L,\lambda}^+(\epsilon)
\, . 
\label{mfd}
\end{align}
Equations (\ref{mmkgf}) and (\ref{mfd}) are equivalent to Eqs.~(35) and (36) in Ref.~\onlinecite{YU2015}. 
The particle number counting field $\chi$ shifts the discretized counting field by $-\chi/M$, 
\begin{align}
\lambda_\ell
\to \lambda_\ell -\chi/M
\, , 
\label{shift}
\end{align}
see the lhs of Eq.~(\ref{dft}).

The multi-contour Keldysh Green function for an electron in the subsystem $B$ is defined as (see Eq.~(37) in Ref.~\onlinecite{YU2015}), 
\begin{align}
g_{R k}^{ms,m's'}(t,t')
=
-i 
\left \langle 
\hat{T}_C
\hat{a}_{R k}(t_{ms})_I
\hat{a}_{R k}^\dagger(t_{m s'}')_I
\right \rangle_M \, . 
\end{align}
The $2 \times 2$ sub-matrix of the $2M \times 2M$ Keldysh Green function matrix is $\left[ {\mathbf g}_{R k}(t,t') \right]_{m,m'} = {\mathbf g}_{R k}(t,t') \delta_{m,m'}$, where
(see Eq.~(38) in Ref.~\onlinecite{YU2015})
\begin{widetext}
\begin{align}
{\mathbf g}_{R k}(t,t')
=
-i e^{-i \epsilon_{Rk} (t-t')}
\left[
\begin{array}{cc}
f_R^-(\epsilon_{Rk}) \theta(t-t') - f_R^+(\epsilon_{Rk}) \theta(t'-t) & 
f_R^+(\epsilon_{Rk}) \\ 
-f_R^-(\epsilon_{Rk}) & 
f_R^-(\epsilon_{Rk}) \theta(t'-t) - f_R^+(\epsilon_{Rk}) \theta(t-t')
\end{array}
\right]
\, . 
\end{align}
\end{widetext}

\section{R\'enyi entropy of the spinless resonant level model}
\label{srlm}

The modified R\'enyi entropy (\ref{renyiglobal}) can be calculated by performing the linked cluster expansion of the Keldysh partition function (\ref{kpf}) using the multi-contour Keldysh Green function. 
The detailed calculations are almost the same as those in Sec. V. A in Ref.~\onlinecite{YU2015}. 
The result is Eq.~(42) in Ref.~\onlinecite{YU2015} (without the spin index) modified with the shift of the discretized counting field (\ref{shift}); 
\begin{align}
\ln 
\frac{S_{M}(\chi)}{s_{M}(\chi)}
=
\sum_{\ell =0}^{M-1}
{\mathcal W}_{\tau}(\lambda_\ell-\chi/M)
\, . 
\label{maires}
\end{align}
In the limit of long measurement time $\tau$, we consider the scaled cumulant generating function (Eqs.~(53) and (54) in Ref.~\onlinecite{YU2015}), 
\begin{align}
{\mathcal F}_{G}(\lambda)
=&
\lim_{\tau \to \infty}
\ln {\mathcal W}_{\tau}(\lambda)/\tau
\nonumber \\
=&
\frac{1}{2 \pi}
\int d \omega
\ln 
\frac{ \tilde{f}_L^+(\omega) + \tilde{f}_L^-(\omega) e^{i \lambda} }{ {f}_L^+(\omega) + {f}_L^-(\omega) e^{i \lambda} }
\, . 
\label{scgffree}
\end{align}
Here we subtracted a trivial constant in order to satisfy the normalization condition ${\mathcal F}_{G}(0)=0$. 
We introduced the effective electron (hole) distribution function 
$ \tilde{f}_L^\pm(\omega) = {\mathcal T}(\omega) f_R^\pm(\omega) + {\mathcal R}(\omega) f_L^\pm(\omega) $, 
where ${\mathcal T}(\omega)$ is the transmission probability and ${\mathcal R}(\omega) = 1-{\mathcal T}(\omega)$ is the reflection probability. 
The explicit form is, 
\begin{align}
{\mathcal T}(\omega)
=
\frac{\Gamma_L \Gamma_R}{(\omega - \epsilon_D)^2 + \Gamma^2/4}
\, ,
\;\;\;\;
\Gamma = \Gamma_L + \Gamma_R
\, . 
\end{align}
The coupling strength between the quantum dot and the lead $r$, $\Gamma_r = 2 \pi \sum_k |J_r|^2 \delta (\omega - \epsilon_{rk})$, 
is assumed to be energy independent. 
In the following, we raise the right chemical potential $\mu = \mu_R - \mu_L>0$, so that electrons are emitted from the right lead, i.e., subsystem $B$. 
In the zero-temperature limit, ${\mathcal F}_{G}$ becomes the scaled cumulant generating function of the number of transmitted particles, i.e., the Levitov-Lesovik formula~\cite{Levitov}, 
\begin{align}
{\mathcal F}_{G}(\lambda)
=
\frac{1}{2 \pi}
\int_{\mu_L}^{\mu_R} d \omega
\ln 
\left[
1+
{\mathcal T}(\omega)
(e^{-i \lambda}-1)
\right]
\, . 
\end{align}
By substituting it into Eq.~(\ref{maires}), we obtain
\begin{align}
\ln 
S_{M}(\chi)
\approx &
\tau
\int_{\mu_L}^{\mu_R} 
\frac{d \omega}{2 \pi}
\, 
\ln 
\left[ 
{\mathcal T}(\omega)^M e^{i \chi}
+
{\mathcal R}(\omega)^M
\right]
\nonumber \\
&+i \chi N_{A,0}
\, , 
\label{igfrlm}
\end{align}
where the second line of rhs is the modified R\'enyi entropy of initial decoupled systems (\ref{modrendec}).

\subsection{Energy-independent transmission}

In the generalized wide-band limit~\cite{HNB}, 
$\Gamma \gg \mu$
or
$|\epsilon_D - \mu_r| \gg \Gamma, \mu$
, 
the transmission and reflection probabilities are energy independent. 
The modified R\'enyi entropy ~(\ref{igfrlm}) reads,
\begin{align}
S_{M}(\chi)
\approx
\left(
{\mathcal T}^M e^{i \chi}
+
{\mathcal R}^M
\right)^{N_{\rm att}}
e^{i \chi N_{A,0}}
\, . 
\label{igfeneind}
\end{align}
Here 
$N_{\rm att}=\tau \mu/(2 \pi)$ 
is the number of particles injected from the right lead. 
The R\'enyi entropy (\ref{renyi}) is obtained by the inverse Fourier transform of Eq.~(\ref{igfeneind}). 
For a non-negative integer $N_{\rm att}$, we obtain
\begin{align}
S_{M}(N_A)
\approx
\binom{N_{\rm att}}{\Delta N_A}
\left(
{\mathcal T}^{\Delta N_A} {\mathcal R}^{N_{\rm att}-\Delta N_A}
\right)^M
\, , 
\label{renyinwbl}
\end{align}
where $\Delta N_A=N_A-N_{A,0}=0, \cdots, N_{\rm att}$ is the number of transmitted particles. 
The probability of finding $N_A$ particles (\ref{ren2pro}) obeys the binomial distribution, 
\begin{align}
P(N_A)
=
S_1(N_A)
\approx
\binom{N_{\rm att}}{\Delta N_A}
{\mathcal T}^{\Delta N_A} {\mathcal R}^{N_{\rm att}-\Delta N_A}
\, . 
\label{binom}
\end{align}
The `Jarzynski equality' (\ref{jarequ}) results in the binomial coefficient;  
\begin{align}
S_0(N_A) \approx \binom{N_{\rm att}}{\Delta N_A} \, . 
\label{Jar_eq}
\end{align}

The joint probability distribution (\ref{jpdfin}) is obtained from Eq.~(\ref{renyinwbl}) by the analytic continuation $M \to 1-i \xi$ and the inverse Fourier transform; 
\begin{align}
P(I_A^\prime,N_A)
\approx&
P(N_A)
\delta
\left(
I_A^\prime
+
\ln
{\mathcal T}^{\Delta N_A} {\mathcal R}^{N_{\rm att}-\Delta N_A}
\right)
\, . 
\label{P_IN}
\end{align}
By substituting it into Eq.~(\ref{pdfci}), we obtain the probability distribution of the conditional self-information, 
\begin{align}
P(J) \approx \sum_{\Delta N_A=0}^{N_{\rm att}} 
P(N_A) \delta \left( J - \ln S_0(N_A) \right) \, , 
\label{pdfj}
\end{align}
which implies that the conditional self-information measures the size of available states in the Fock subspace containing $N_A$ particles~(\ref{jarequ}).

\subsubsection{Time dependence of the accessible entanglement entropy}

\begin{figure}[ht]
\includegraphics[width=0.7 \columnwidth]{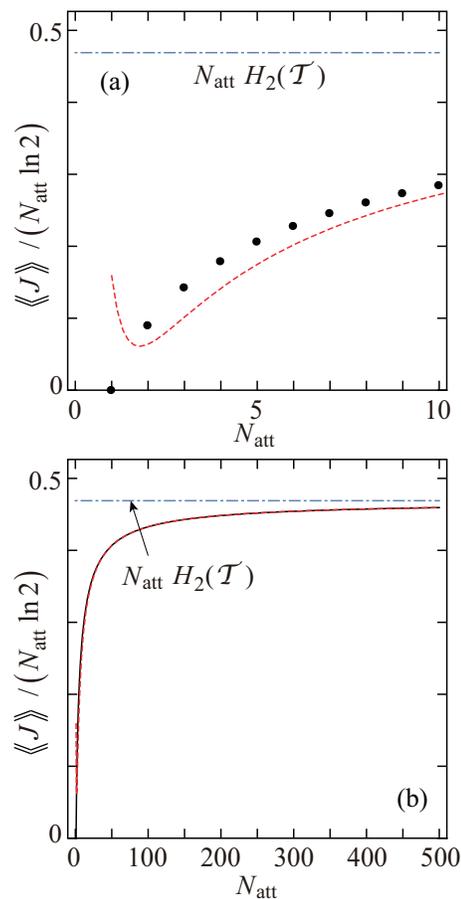}
\caption{
The time dependence of the conditional entropy, i.e., the accessible entanglement entropy, for ${\mathcal T}=0.1$. 
Panels (a) and (b) show results in the short-and long-time regimes, respectively. 
The dashed lines represent the approximation in the long-time regime (\ref{gaussapprox}). 
The dashed line in panel (b) overlaps with the numerical result. 
The dot-dashed lines indicate the full entanglement entropy. 
}
\label{tvsj}
\end{figure}

Figures \ref{tvsj} (a) and (b) show the conditional entropy, i.e., the accessible entanglement entropy, as a function of the measurement time $N_{\rm att} = \tau \mu/(2 \pi)$. 
At $\tau=0$, i.e., $N_{\rm att}=0$, we have trivially $P(J)=\delta(J)$, where we used $0!=1$. 
At $\tau=\pi/\mu$, i.e., when only one electron is injected $N_{\rm att}=1$, we obtain, 
\begin{subequations}
\begin{align}
P(J) \approx {\mathcal R} \delta(J) + {\mathcal T} \delta(J) = \delta(J). 
\end{align}
Therefore, a single particle cannot create the accessible entanglement entropy $ \langle \! \langle  J \rangle \! \rangle =0$ [Fig.~\ref{tvsj} (a)], as is consistent with the local-particle number superselection rule~\cite{Beenakker,Klich2,Wiseman2003,Dowling2006}. 
At $N_{\rm att}=2$, when two electrons participate, the probability distribution is, 
\begin{align}
P(J)
\approx
(
{\mathcal T}^2 + {\mathcal R}^2 
)
\delta(J)
+
2
{\mathcal T} {\mathcal R} \delta(J-\ln 2)
\, . 
\label{pdfjn2}
\end{align}
The accessible entanglement entropy is, 
$ \langle \! \langle  J \rangle \! \rangle = 2 {\mathcal T} {\mathcal R} \ln 2 $
[Fig.~\ref{tvsj} (a)], 
which is consistent with the previous theory~\cite{Beenakker}. 
The second term of Eq.~(\ref{pdfjn2}) is associated with a single EPR pair. 
Since we consider spinless fermions, the entanglement is attributed to the orbital degree of freedom. 

When three particles participate, $N_{\rm att}=3$, we obtain, 
\begin{align}
P(J) \approx ({\mathcal T}^3 +{\mathcal R}^3 )\delta(J) + 3 {\mathcal T} {\mathcal R} \delta(J-\ln 3). 
\end{align}
\end{subequations}
The accessible entanglement entropy is, 
$\langle \! \langle J \rangle \! \rangle = 3 {\mathcal T} {\mathcal R} \ln 3$. 
In the short-time regime, the accessible entanglement entropy depends nonlinearly on time [Fig.~\ref{tvsj} (a)]. 
The result is different from that with the previous theory~\cite{Beenakker}, which predicted that accessible entanglement entropy increases linearly as a function of measurement time. 
The difference arises because we consider the measurement of the total particle number in subsystem $A$, while the previous theory~\cite{Beenakker} considered the measurement of the particle number of each energy level in subsystem $A$. 

In the above discussions, for simplicity, we applied the scaled cumulant generating function (\ref{scgffree}) to the short-time regime. 
We neglected corrections due to finite measurement time~\cite{Muzykanskii2003}, which induces fluctuations in the number of injected particles $N_{\rm att}$.

In Fig.~\ref{tvsj} (b) we observe that, as the measurement time increases, the conditional entropy increases and approaches full entanglement entropy
$\langle \! \langle I_A \rangle \! \rangle = N_{\rm att} H_2({\mathcal T})$ 
(dot-dashed line). 
Here the binary entropy is, 
$ H_2(x) = -x \ln x -(1-x) \ln (1-x) $. 
Since the full entanglement entropy is proportional to the measurement time $\tau \propto N_{\rm att}$, from the chain rule (\ref{ententsupsel1}), we conclude that the nonlinear time dependence observed in Fig.~\ref{tvsj} is attributable to the time dependence of the Shannon entropy $H(N_A)$ (\ref{Shannonentropy}) of the binomial distribution (\ref{binom}).

In the limit of long measurement time, when many particles participate, $N_{\rm att} \gg 1$, we perform the Gaussian approximation, 
$ \ln S_1(\chi) \approx i \chi (N_{A,0} +N_{\rm att} {\mathcal T} ) + (i \chi)^2 N_{\rm att} {\mathcal T} {\mathcal R}/2! $
and obtain the Shannon entropy, 
$ H(N_A) \approx \ln \sqrt{ 2 \pi e N_{\rm att} {\mathcal T} {\mathcal R} } $. 
This is the $\ln \tau$ subleading correction derived in Ref.~\onlinecite{Klich2}. 
For $N_{\rm att} \gg 1$, the electron number is most probably $\langle \! \langle N_A \rangle \! \rangle = N_{A,0} +N_{\rm att} {\mathcal T}$. 
The remaining uncertainty, which is the origin of this Shannon entropy, is attributable to the width of the distribution $\langle \! \langle N_A^2 \rangle \! \rangle = N_{\rm att} {\mathcal T} {\mathcal R}$. 
The dotted lines in Fig.~\ref{tvsj} represent the approximation 
\begin{align}
\langle \! \langle J \rangle \! \rangle \approx N_{\rm att} H_2({\mathcal T}) - \ln \sqrt{ 2 \pi e N_{\rm att} {\mathcal T} {\mathcal R} } \, . 
\label{gaussapprox}
\end{align} 
The dotted line fits the result in the long-time regime [panel (b)]. 
Deviations are observed in the short-time regime [panel (a)].

\subsubsection{Probability distribution of the conditional self-information}

Here, we analyze the distribution of the conditional self-information in the limit of a long measurement time, $\tau \propto N_{\rm att} \gg 1$. 
In this case, the inverse Fourier transform of the information generating function (\ref{rigf}) can be done within the saddlepoint approximation, i.e., the Legendre-Fenchel transform~\cite{Touchette}; 
\begin{align}
P(J)
=&
\int
\frac{d \xi}{2 \pi}
e^{-i \xi J}
R_{1-i \xi}
\nonumber \\
\approx&
\exp
\left[
N_{\rm att}
\min_{i \xi \in \mathbb{R}}
\left(
\ln R_{1-i \xi}
-
i \xi J 
\right)
\right]
\, . 
\label{lfrp}
\end{align}
Therefore, we can assume that $\xi$ is a pure imaginary number. 
In the rest of this section, we regard $i \xi=1-M$ as a real number. 
By substituting Eq.~(\ref{renyinwbl}) into Eq.~(\ref{rigf}), we obtain the information generating function, 
\begin{align}
R_M
\approx&
\sum_{\Delta {N_A}=0}^{N_{\rm att}}
{\mathcal T}^{\Delta N_A} {\mathcal R}^{N_{\rm att}-\Delta N_A}
\binom{N_{\rm att}}{\Delta N_A}^{2-M}
\, .  
\end{align}
 
\begin{figure}[ht]
\includegraphics[width=0.7 \columnwidth]{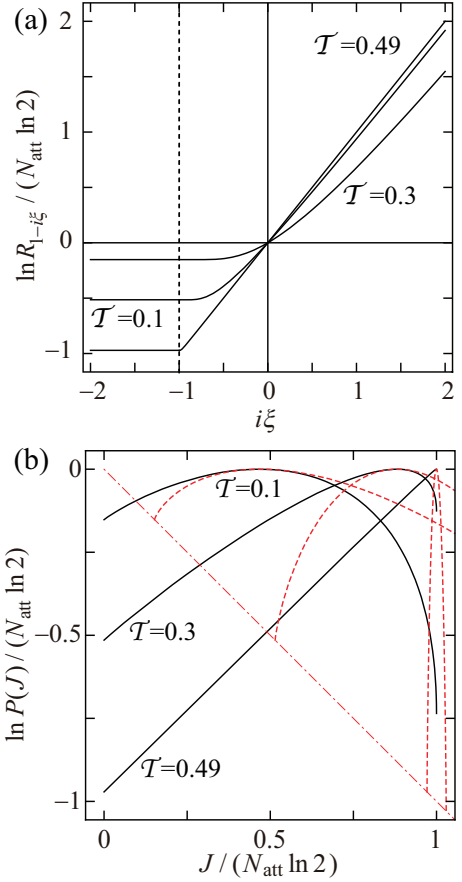}
\caption{
The information generating function (a) and the probability distribution of the conditional self-information (b). 
The dashed lines in panel (b) indicate the probability distribution of the self-information $P(I_A)$ taken from Fig.~5 (b) of Ref.~\onlinecite{YU2015}. 
}
\label{rpconent}
\end{figure}

For $N_{\rm att} \gg1$, we can replace the summation with the integral and utilize Stirling's formula. 
Then we obtain, 
\begin{subequations}
\begin{align}
R_M
\approx &
\int_0^{N_{\rm att}} dn
\exp
\left[
- N_{\rm att} \, D_M(n/N_{\rm att})
\right]
\, ,
\\
D_M(p)
=&
-p \ln {\mathcal T} - (1-p) \ln {\mathcal R}
+
(M-2)
H_2(p)
\, . 
\label{omeM}
\end{align}
\end{subequations}
Within the saddlepoint approximation we obtain, 
\begin{align}
\frac{\ln R_M}{N_{\rm att}}
\approx &
-
\min_{0 \leq p \leq 1} \, D_M(p)
\nonumber \\
=&
\left \{
\begin{array}{cc}
\ln
\left( {\mathcal T}^{1/(2-M)} + {\mathcal R}^{1/(2-M)} \right)^{2-M}
& (M \leq 2) \\
\min ( 
\ln {\mathcal T},
\ln {\mathcal R}
) & (M>2) 
\end{array}
\right.
\, . 
\label{relrenbin}
\end{align}
The first and second cumulants of the conditional self-information calculated by the derivative~(\ref{jcum}) 
are compatible with those of the full entanglement entropy; 
$ \langle \! \langle J \rangle \! \rangle = \langle \! \langle {I_A^\prime} \rangle \! \rangle = N_{\rm att} H_2({\mathcal T})$
and 
$ \langle \! \langle J^2 \rangle \! \rangle = \langle \! \langle {I_A^\prime}^2 \rangle \! \rangle
=
N_{\rm att}
{\mathcal T} 
{\mathcal R}
[\ln ({\mathcal R}/{\mathcal T})]^2
$. 
However, the skewness is different, 
$ \langle \! \langle J^3 \rangle \! \rangle = \langle \! \langle {I_A^\prime}^3 \rangle \! \rangle -3 \langle \! \langle J^2 \rangle \! \rangle $, 
where 
$ \langle \! \langle {I_A^\prime}^3 \rangle \! \rangle = \langle \! \langle {I_A^\prime}^2 \rangle \! \rangle ({\mathcal R}-{\mathcal T}) \ln ({\mathcal R}/{\mathcal T}) $. 
Figures~\ref{rpconent} (a) and (b) show the information generating function and corresponding probability distribution obtained from the Legendre-Fenchel transform~(\ref{lfrp}). 
The Legendre duality~\cite{Touchette} implies that the maximum (minimum) value of fluctuating $J$ is the slope of the logarithm of the information generating function at $i \xi \to \infty $ ($-\infty$) [Fig.~\ref{rpconent} (a)];  
\begin{subequations}
\begin{align}
J_{\rm max}
=&
\lim_{i \xi \to \infty}
\frac{ \ln R_{1-i \xi} }{i \xi}
=
N_{\rm att} \ln2 
\, , 
\label{jmax}
\\
J_{\rm min}
=&
\lim_{i \xi \to -\infty}
\frac{ \ln R_{1-i \xi} }{i \xi}
=
0
\, . 
\label{jmin}
\end{align}
\end{subequations}
The dashed lines in Fig.~\ref{rpconent} (b) depict the probability distributions of the full entanglement entropy $P(I_A)$ taken from Fig.~5 (b) of Ref.~\onlinecite{YU2015}. 
Although the peak positions coincide, the minimum and maximum values differ. 
In Ref.~\onlinecite{YU2015}, we obtained 
${I_A}_{\rm max} = N_{\rm att} \max( -\ln {\mathcal T}, -\ln {\mathcal R})$
and
${I_A}_{\rm min} = N_{\rm att} \min( -\ln {\mathcal T}, -\ln {\mathcal R})$.

The minimum and maximum values of $J$ are also deduced from the probability distribution function (\ref{pdfj}). 
The maximum value $J_{\rm max}$ (\ref{jmax}) corresponding to the maximum $S_0(N_A)$ is obtained when half of the injected electrons are transmitted, $\Delta N_A \approx N_{\rm att}/2$. 
One can check that $-\ln S_0(N_A) = - \ln \binom{N_{\rm att}}{N_{\rm att}/2} \approx {N_{\rm att}} \ln 2$ is the maximum value. 
The probability of finding the maximum value is, 
$ P(J_{\rm max}) \approx \binom{N_{\rm att}}{N_{\rm att}/2} ({\mathcal T} {\mathcal R})^{N_{\rm att}/2} \approx (4 {\mathcal T} {\mathcal R})^{N_{\rm att}/2} $.
The minimum value $J_{\rm min}=0$ (\ref{jmin}) is achieved when all injected electrons are transmitted $\Delta N_A=N_{\rm att}$ or reflected $\Delta N_A=0$. 
One can check that both cases result in the minimum value $-\ln S_0(N_A) = - \ln \binom{N_{\rm att}}{0} = - \ln \binom{N_{\rm att}}{N_{\rm att}} =0$. 
The probability of finding the minimum value is, 
$ P(J_{\rm min}) = {\mathcal T}^{N_{\rm att}} + {\mathcal R}^{N_{\rm att}} $. 
%
The mode corresponds to events when $\Delta {N_A} \approx N_{\rm att} {\mathcal T}$ electrons are transmitted; 
$ J_{\rm mode} \approx \ln \binom{N_{\rm att}}{N_{\rm att} {\mathcal T}} \approx N_{\rm att} H_2 ({\mathcal T})$. 
The probability to find this value is almost 1, 
$ P(J_{\rm mode}) \approx \binom{N_{\rm att}}{N_{\rm att} {\mathcal T}} {\mathcal T}^{N_{\rm att} {\mathcal T}} {\mathcal R}^{N_{\rm att} {\mathcal R}} \approx
1$. 

\section{Joint probability distribution}

\label{edtp}

For the energy-independent transmission probability, second joint cumulants are calculated from Eqs.~(\ref{joicum}) and ~(\ref{igfeneind}) as, 
$\langle \! \langle N_A^2 \rangle \! \rangle = N_{\rm att} {\mathcal T} {\mathcal R}$, 
$\langle \! \langle {I_A^\prime}^2 \rangle \! \rangle = 4 N_{\rm att} {\mathcal T} {\mathcal R} [ \tanh^{-1} ({\mathcal R}-{\mathcal T})]^2$ 
and
$\langle \! \langle I_A^\prime N_A \rangle \! \rangle = 2 N_{\rm att} {\mathcal T} {\mathcal R} \tanh^{-1} ({\mathcal R}-{\mathcal T})$. 
Then the correlation coefficient (\ref{corcoe}) is,  
\begin{align}
r={\rm sgn}({\mathcal R}-{\mathcal T}) \, . 
\label{corcoeeneind}
\end{align}
Since $|r|=1$, there exists a perfect linear correlation between the self information and the particle number. 
The two quantities are negatively (positively) correlated for large (small) transmission probability ${\mathcal T}>{\mathcal R}$ (${\mathcal T}<{\mathcal R}$). 
The perfect linear correlation can be also deduced from Eq.~(\ref{P_IN}). 
The argument of the delta function is zero when two quantities are linearly correlated; $I_A^\prime = \Delta N_A \ln ({\mathcal R}/{\mathcal T}) - N_{\rm att} \ln {\mathcal R}$. 
The correlation coefficient (\ref{corcoeeneind}) implies that when the transmission probability is energy independent, one can determine the self-information and consequently the entanglement entropy by counting the number of electrons. 

The energy dependence of the transmission probability spoils the perfect linear correlation. 
In the following, we will analyze the joint probability distribution for such a case in the limit of long measurement time $\tau \to \infty$. 
We will limit our discussion to the symmetric case; $\Gamma_L=\Gamma_R$ and $\mu_R-\epsilon_D=\epsilon_D-\mu_L=\mu/2$. 
The transmission probability is, 
\begin{align}
{\mathcal T}(\omega)
=
\frac{1}{1+z^2}
\, ,
\;\;\;
z
=
\frac{\omega-\epsilon_D}{\Gamma/2}
\, . 
\label{sc}
\end{align}

\subsection{Small bias voltage: Coherent resonant tunneling}

For a small bias voltage $0<v = \mu/\Gamma < 1$ and for a positive integer $M$, an analytic expression of the modified information generating function can be obtained. 
By noting that $|e^{i \chi}|=1$, we first expand the integrand of Eq.~(\ref{igfrlm}) in powers of $z$ (\ref{sc}) and then perform the integral. 
The result is, 
\begin{align}
\ln S_M(\chi)
=&
N_{\rm att}
(i \chi+F_M(\chi)-M F_1(0))
+
i \chi N_{A,0}
\, ,
\nonumber \\
F_M(\chi)
=&
\ln 
(1+v^{2M} e^{-i \chi}) 
\nonumber \\ &
+
\Phi
(-v^{2M} e^{-i \chi},1,1/(2M))
\, , 
\label{igfsv}
\end{align}
where $\Phi$ is the Hurwitz-Lerch zeta function~\cite{NISTbook}; 
\begin{align}
\Phi(z,s,a)
=
\sum_{k=0}
\frac{z^k}{(k+a)^s}
\, , 
\;\;
(
|z|<1 \, , 
a \neq 0, -1,-2,\cdots  \, ) 
\, . 
\end{align}
The scaled cumulant generating function of the particle number is obtained from Eq.~(\ref{igfsv}) by setting $M=1$. 
Up to the third order in $v$, and noting that $N_{\rm att}= v \tau \Gamma/(2 \pi)$ is also proportional to $v$, the result is, 
\begin{align}
\ln S_1(\chi)
\approx
i \chi
( N_{\rm att} + N_{A,0} )
+
N_{\rm att}
\frac{v^2}{3}
(e^{-i \chi}-1)
\, .
\end{align}
The first term is attributable to the bulk electrons in the left lead and electrons injected from the right lead without scattering. 
The second term is caused by uncorrelated backscattering events. 
The strength of the backscattering is proportional to $v^3$, which is a property of the Fermi liquid~\cite{Gogolin2006,Sakano2011,Sakano2012}. 
The first and second cumulants are analytic in $v$ as, 
\begin{subequations}
\begin{align}
\langle \! \langle \Delta N_A \rangle \! \rangle
=&
N_{\rm att}
\left(1-\frac{v^2}{3} \right)
\, ,
\label{c1nsmall}
\\
\langle \! \langle N_A^2 \rangle \! \rangle
=&
N_{\rm att}
\frac{v^2}{3}
\, . 
\end{align}
\end{subequations}

\begin{figure}[ht]
\includegraphics[width=0.7 \columnwidth]{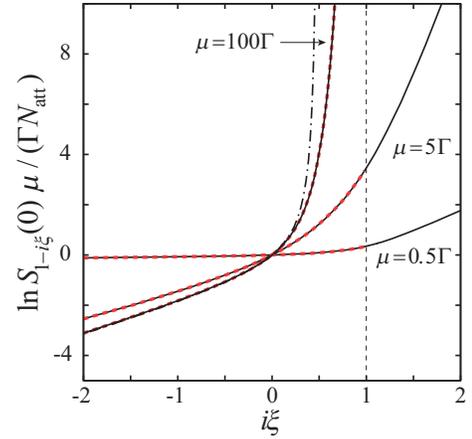}
\caption{
The information generating function for various bias voltages ($v=\mu/\Gamma=0.5,5,100$). 
Dotted lines show the analytic expression for $v<1$ (\ref{igfsmallv}) applicable to the range $i \xi <1$. 
They almost overlap with the solid lines. 
The dot-dashed line indicates the analytic expression in the limit of $v \to \infty$ (\ref{ifglargev}).  
}
\label{ldf}
\end{figure}

The information generating function (\ref{renyif}) is obtained from Eq.~(\ref{igfsv}) by setting $\chi=0$, 
\begin{align}
\ln S_M(0)
=&
N_{\rm att}
\left [
\ln (1+v^{2M})/(1+v^2)^M
\right. 
\nonumber \\
&
\left. 
+
\Phi(-v^{2M},1,1/(2M))
-2M (\tan^{-1} v)/v
\right]
\, ,
\label{igfsmallv}
\end{align}
and by analytically continuing $M$ to a real number$1 - i \xi$. 
The range is limited to $i \xi=1-M<1$, since the Hurwitz-Lerch zeta function diverges at $i \xi=1-M=\infty,1+1/2,1+1/4, \cdots, 1+0$. 
We note that it still satisfies the `Jarzynski equality' (\ref{jarequ2015}), 
$ \lim_{M \to +0} \ln S_M(0) = N_{\rm att} \ln 2 $. 
The solid lines in Fig.~\ref{ldf} are the information generating function (\ref{igfrlm}) for small ($v=\mu/\Gamma=0.5$) and large ($v=5$ and $v=100$) bias voltages. 
They are well fitted by the analytic expression (\ref{igfsmallv}) depicted by dotted lines even for large bias voltages $v=\mu/\Gamma>1$. 

The first and second cumulants of the self-information and the covariance are nonanalytic in $v$, 
\begin{subequations}
\begin{align}
\langle \! \langle I_A^\prime \rangle \! \rangle
\approx &
N_{\rm att}
\frac{v^2}{9}
\left(5-6 \ln v \right)
\, ,
\label{c1ismall}
\\
\langle \! \langle {I_A^\prime}^2 \rangle \! \rangle
\approx&
N_{\rm att}
\frac{4 v^2}{27}
\{ 1 + [\ln (e v^{-3})]^2 \}
\, , 
\\
\langle \! \langle I^\prime N_A \rangle \! \rangle
\approx&
-
N_{\rm att}
\frac{2 v^2}{9}
\ln (e v^{-3})
\,  , 
\end{align}
\end{subequations}
They vanish in the limit of zero bias voltage $v \to +0$. 
Since the covariance $\langle \! \langle I^\prime N_A \rangle \! \rangle$ is negative, the correlation coefficient (\ref{corcoe}) is also negative, 
\begin{align}
r
\approx 
-
\frac{\ln (e v^{-3})}
{\sqrt{1+[\ln (e v^{-3})]^2}}
\approx
-1
+
\frac{1}{2 [\ln (e v^{-3})]^2}
\,  .
\end{align}
In the limit of small bias voltage $v \to +0$, it approaches $-1$, since the transmission probabilities of electrons inside the Fermi window $\mu_L <\omega < \mu_R$ are almost 1; Equation~(\ref{corcoeeneind}) implies that, for the perfect transmission, the correlation coefficient is $-1$. 
A finite bias voltage induces a nonvanishing reflection probability, which results in $(\ln v)^{-2}$ correction and spoils the perfect linear correlation.

\subsection{Large bias voltage: Incoherent sequential tunneling}

For $0<e^{i \chi} \in \mathbb{R}$, the integral in Eq.~(\ref{igfrlm}) can be done in the limit of large bias voltage $v \to \infty$, 
\begin{align}
\ln 
S_{M}(\chi)
\approx
\frac{\tau \Gamma}{2}
\left[
e^{i \chi/(2 M)}
\csc \left( \frac{\pi}{2M} \right)
-
M
\right]
+
i \chi N_{A,0}
\, . 
\label{cgfnilargev}
\end{align}
The scaled cumulant generating function of the particle number is then, 
\begin{subequations}
\begin{align}
\ln S_1(\chi) 
\approx &
i \chi N_{A,0}
+
\tau {\mathcal F}_{\rm seq}(-\chi)
\, ,
\\
{\mathcal F}_{\rm seq}(\lambda)
= &
\frac{\Gamma}{2}
(e^{-i \lambda/2}-1)
\, . 
\label{cgfnlargev}
\end{align}
\end{subequations}
Here Eq.~(\ref{cgfnlargev}) reproduces the scaled cumulant generating function of the incoherent sequential tunneling, which was derived based on the master equation approach of full counting statistics~\cite{Bagrets2003}. 
One can check that Eq.~(\ref{cgfnilargev}), except for the bulk contribution $i \chi N_{A,0}$, can be obtained by substituting Eq.~(\ref{cgfnlargev}) into Eq.~(\ref{maires}); 
\begin{align}
\sum_{\ell=0}^{M-1}
\tau 
{\mathcal F}_{\rm seq}(\lambda_\ell - \chi/M)
=
\frac{\tau \Gamma}{2}
\left[
e^{i \chi/(2 M)}
\csc \left( \frac{\pi}{2M} \right)
-
M
\right]
\, . 
\end{align}
This implies that even in the incoherent sequential tunneling regime, the entanglement entropy is attributable to the particle fluctuations at the boundary. 

The information generating function (\ref{renyif}) is derived from Eq.~(\ref{cgfnilargev}); 
\begin{align}
\ln 
S_{M}(0)
\approx 
\frac{\tau \Gamma}{2}
\left[
\csc \left( \frac{\pi}{2M} \right)
-
M
\right]
\, . 
\label{ifglargev}
\end{align}
It diverges at $M=1-i \xi =1/2$ [dot-dashed line in Fig.~\ref{ldf}]. 
The expression implies that all cumulants are proportional to the coupling strength $\Gamma$. 
The first cumulants and second joint cumulants are, 
\begin{subequations}
\begin{align}
\langle \! \langle I_A^\prime \rangle \! \rangle/2
=&
\langle \! \langle \Delta N_A \rangle \! \rangle
=
\tau \Gamma/4
\, .
\label{aveinc}
\\
(2/\pi^2) \, \langle \! \langle {I_A^\prime}^2 \rangle \! \rangle 
=&
2 \langle \! \langle {N_A}^2 \rangle \! \rangle
=
\langle \! \langle  I_A^\prime {N_A} \rangle \! \rangle
=
\langle \! \langle \Delta N_A \rangle \! \rangle
\,  . 
\end{align}
\end{subequations}
Then the correlation coefficient (\ref{corcoe}) is positive and is independent of $\Gamma$, 
\begin{align}
r
=
\frac{2}{\pi}
\approx
0.636
\,  . 
\end{align}

\subsection{Contour plot}

\begin{figure}[ht]
\includegraphics[width=0.6 \columnwidth]{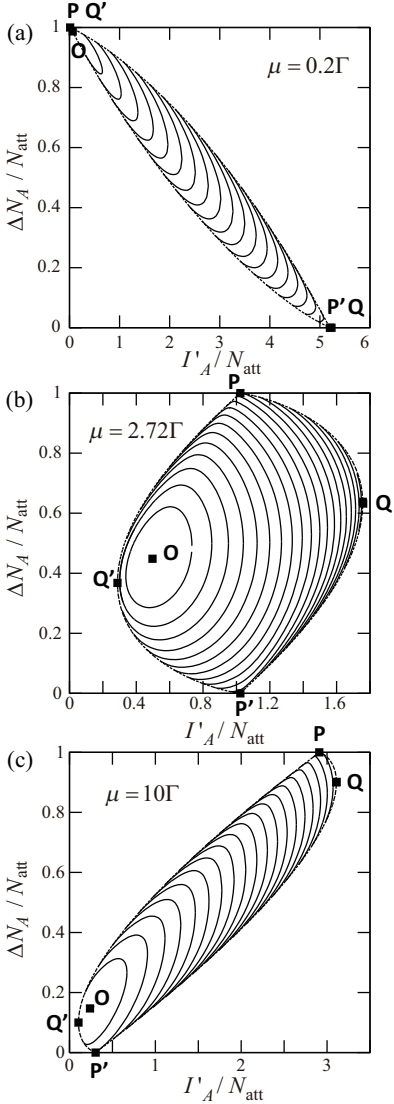}
\caption{
Contour plots of the logarithm of the joint probability distribution $\ln P(I_A^\prime,N_A)$ for (a) a small bias voltage $v=\mu/\Gamma=0.2$, (b) an intermediate bias voltage $v=2.72$, and (c) a large bias voltage $v=10$. 
In each panel, the support, the region with positive probabilities, is surrounded by a dashed line. 
Squires ${\mathbf P}$ and ${\mathbf P}'$ indicate boundary points of support with maximum and minimum $N_A$, respectively. 
Squires ${\mathbf Q}$ and ${\mathbf Q}'$ indicate boundary points of support with maximum and minimum $I_A^\prime$, respectively. 
Minimum values of $P(I_{\rm A}^\prime,N_A)$ are ${\rm min}_{I_{\rm A}^\prime,N_A} \ln P(I_{\rm A}^\prime,N_A)/N_{\rm att}=-5.20$ (a),$-1.74$, (b) and $-3.11$ (c).
}
\label{cp}
\end{figure}

The three panels in Fig.~\ref{cp} show contour plots of the joint probability distribution of self-information and the particle number obtained within the Legendre-Fenchel transform~\cite{Touchette} of the information generating function derived from the modified R\'enyi entropy (\ref{igfrlm}); 
\begin{align}
P(I_A^\prime,N_A)
=&
\int \frac{d \xi}{2 \pi}
\int_{-\pi}^\pi \frac{d \chi}{2 \pi}
e^{-i \xi I_A^\prime - i \chi N_A}
S_{1-i \xi}(\chi)
\nonumber \\
\approx &
\exp
\left[
\min_{i \xi,i \chi \in \mathbb{R}}
\left(
-i \xi I_A^\prime - i \chi N_A
+
\ln S_{1-i \xi}(\chi)
\right)
\right]
\,  . 
\nonumber
\end{align}
In each panel, the peak is at ${\mathbf O}=(\langle \! \langle I_A^\prime \rangle \! \rangle, \langle \! \langle N_A \rangle \! \rangle)$. 
We observe that fluctuations in $N_A$ and $I_A^\prime$ are bounded. 
The support, which is the region in the $(I_A^\prime,N_A)$ plane with positive probabilities,~\cite{Hogg} is surrounded by a dashed line. 

Panel (a) shows a plot for a small bias voltage $v=\mu/\Gamma=0.2$. 
We observe a negative linear correlation. 
Let us derive the maximum of $N_A$. 
For $i \chi \to \infty$, the modified R\'enyi entropy behaves as, 
\begin{align}
\ln S_M(\chi)
\approx &
\frac{\tau}{2 \pi}
\int_{\mu_L}^{\mu_R}
d \omega
\ln {\mathcal T}(\omega)^M e^{i \chi}
+i \chi N_{A,0}
\nonumber \\
=&
N_{\rm att}(i \chi - M F_1(0) )
+i \chi N_{A,0}
\, . 
\end{align}
Then the Legendre duality~\cite{Touchette} implies that a boundary point of support with maximum $N_A$ is, 
\begin{align}
{\mathbf P}
=&
\lim_{i \chi \to + \infty}
(
\partial_{i \xi} \ln S_{1-i \xi}(\chi)
,
\partial_{i \chi} \ln S_{1-i \xi}(\chi)
)
\nonumber \\
=&
(N_{\rm att} F_1(0),N_{\rm att}+N_{A,0})
\, . 
\end{align}
A boundary point of support with minimum $N_A$ is derived from the R\'enyi entropy for $i \chi \to - \infty$, 
\begin{align}
\ln S_M(\lambda)
\approx&
\frac{\tau}{2 \pi}
\int_{\mu_L}^{\mu_R}
d \omega
\ln {\mathcal R}(\omega)^M
+
i \chi N_{A,0}
\nonumber \\
=&
-
N_{\rm att}
M 
[F_1(0)- \ln (v/e)^2]
+
i \chi N_{A,0}
\, ,
\end{align}
and the Legendre duality, 
\begin{align}
{\mathbf P}'
=&
\lim_{i \chi \to - \infty}
(
\partial_{i \xi} \ln S_{1-i \xi}(\chi)
,
\partial_{i \chi} \ln S_{1-i \xi}(\chi)
)
\nonumber \\
=&
(N_{\rm att} (F_1(0)-\ln(v/e)^2),N_{A,0})
\, . 
\end{align}
The point ${\mathbf P}$ (${\mathbf P}'$) corresponds to a rare event when all injected $N_{\rm att}$ electrons are transmitted (reflected). 
For a small bias voltage $v<1$, the transmission probability of an electron inside the Fermi window is larger than the reflection probability ${\mathcal T}(\omega)>{\mathcal R}(\omega)$. 
Therefore, at the point ${\mathbf P}$ (${\mathbf P}'$), electrons carry minimum (maximum) self-information $I_A^\prime$, as shown in Fig.~\ref{cp} (a). 
In Appendix \ref{bps}, we calculate a boundary point of support with maximum (minimum) $I_A'$, ${\mathbf Q}$ (${\mathbf Q}'$) and checked that ${\mathbf Q}={\mathbf P}'$ (${\mathbf Q}'={\mathbf P}$).

For a large bias voltage $v>1$, electrons with energy $|\omega-\epsilon_D| > \Gamma/2$ also participate in transmission processes. 
Since the reflection probability of such electrons is larger than the transmission probability ${\mathcal T}(\omega)<{\mathcal R}(\omega)$, the event with $N_{\rm att}$ electron transmission (reflection) does not necessarily carry the minimum (maximum) self-information. 
Figures \ref{cp} (b) and (c) are joint probability distributions for $v>1$. 
The boundary point with maximum (minimum) $N_A$, ${\mathbf P}$ (${\mathbf P}'$), does not coincide with the boundary point with minimum (maximum) $I_A^\prime$, ${\mathbf Q}'$ (${\mathbf Q}$) [Appendix \ref{bps}]. 
For a large bias voltage, we observe a positive linear correlation [Figs.~\ref{cp} (c)].

\section{Probability distribution of efficiency}
\label{peltier}

It would be interesting to consider the similarity between information entropy and thermodynamic entropy. 
Here, we examine an analogy to a thermoelectric effect, the Peltier effect. 
By raising the right chemical potential relative to the left chemical potential by $\mu$, $n$ electrons move from the right lead to the left lead. 
Since electrons also carry the heat $q$ from the right lead to the left lead, our setup works as a heater. 
Its efficiency is characterized by the coefficient of performance (COP)~\cite{Okada2017}; 
$\phi = q/(n \mu)$. 
When the measurement time $\tau$ is short, both $n$ and $q$ fluctuate and thus the COP also fluctuates. 
This problem has been discussed recently~\cite{Verley1,Verley2,Polettini,Jiang,Okada2017,Proesmans2016} in the context of stochastic thermodynamics~\cite{Seifert}. 
It was demonstrated that the Carnot limit corresponds to the rarest event~\cite{Verley1,Verley2,Polettini}. 
In our context, the number of transmitted electrons is $n = \Delta N_A$, and we want to relate the thermodynamic entropy $\beta q$ to the self-information $I_A^\prime$. 
The corresponding (dimensionless) COP $\eta \equiv \beta \mu \phi$ may be, 
\begin{align}
\eta=\frac{I_A^\prime}{\Delta N_A}
\, .
\label{dimlescop} 
\end{align}
It measures the information content carried by a single electron. 

At the steady state, which is achieved in the limit of long measurement time, the average COP would be the average self-information divided by the average number of transmitted electrons $\langle \! \langle \eta \rangle \! \rangle = \langle \! \langle I_A^\prime \rangle \! \rangle/\langle \! \langle \Delta N_A \rangle \! \rangle$. 
At a short measurement time, the COP fluctuates and the (unnormalized) probability distribution of COP may be expressed by using the joint probability distribution (\ref{jpdfin}) as~\cite{Okada2017}, 
\begin{align}
P(\eta)
=&
\int d I_A^\prime
\sum_{N_A \neq N_{A,0}}
P(I_A^\prime,N_A)
\, 
\delta(\eta-I_A^\prime/\Delta N_A)
\nonumber \\
=&
\sum_{\Delta N_A \neq 0}
|\Delta N_A|
P_\tau(\eta \Delta N_A,N_A)
\, . 
\label{pdfcop}
\end{align}

For energy-independent transmission probability, there is a perfect linear correlation between $I_A^\prime$ and $N_A$, see Eq.~(\ref{corcoeeneind}). 
By substituting Eq.~(\ref{P_IN}) into Eq.~(\ref{pdfcop}), we obtain, 
\begin{align}
P(\eta)
=
\sum_{\Delta N_A=1}^{N_{\rm att}}
P(N_A)
\delta \!
\left(
\eta 
+
\ln
{\mathcal T} {\mathcal R}^{N_{\rm att}/\Delta N_A-1}
\right)
\, , 
\end{align}
where $P(N_A)$ is the binomial distribution function~(\ref{binom}). 
We observe that the minimum COP, $\eta_{\rm min} = - \ln {\mathcal T}$, is achieved when all electrons transmit $\Delta N_A=N_{\rm att}$. 
The maximum COP, $\eta_{\rm max} = - \ln {\mathcal T} {\mathcal R}^{N_{\rm att}-1}$, is realized when only one electron transmits $\Delta N_A=1$. 
The probabilities to find these rare events are $P (\eta_{\rm min}) = {\mathcal T}^{N_{\rm att}}$ and $P (\eta_{\rm max}) = N_{\rm att} {\mathcal T} {\mathcal R}^{N_{\rm att}-1}$. 

For long measurement time, $N_{\rm att} \gg 1$, the probability distribution, is calculated as~\cite{Verley1,Verley2}, 
\begin{align}
\ln P(\eta)
=&
\max_{\Delta N_A}
[
\ln P(\eta \Delta N_A,\Delta N_A + N_{A,0})
]
\nonumber \\
=&
\max_{\Delta N_A}
\biggl \{
\min_{i \xi,i \chi}
[
\ln S_{1-i \xi}(\chi)
-
i (\xi \eta + \chi) \Delta N_A 
\nonumber \\
&-
i \chi N_{A,0}
]
\biggl \}
=
\min_{i \xi}
\left(
\ln 
\frac{S_{1-i \xi}(-\xi \eta)}{s_{1-i \xi}(-\xi \eta)}
\right)
\, . 
\label{pdfVerley}
\end{align}
By plugging Eqs.~(\ref{igfeneind}) and (\ref{modrendec}) into the above equation, we find the solution, 
\begin{align}
\ln P(\eta)
=
-
N_{\rm att}
D_1(p^*)
\, , 
\;\;\;\;
p^*
=
\frac{ -\ln {\mathcal R} }{\ln ({\mathcal T}e^\eta / {\mathcal R})}
\, , 
\label{relent}
\end{align}
for $\eta > \eta_{\rm min}$. 
Here $D_{1}$ is defined in Eq.~(\ref{omeM}) and is the relative entropy, the Kullback-Leibler divergence, of $p=\{ p^*,1-p^* \}$ with respect to $q=\{ {\mathcal T}, {\mathcal R} \}$; 
$D_1(p^*) = p^* \ln (p^*/{\mathcal T}) + (1-p^*) \ln [(1-p^*)/{\mathcal R}]$. 
The mode minimizing $D_1(p^*)$ is realized at $p^*={\mathcal T}$. 
The mode is equal to the average at the steady state, 
\begin{align}
\eta_{\rm mode}
=
\frac{
-{\mathcal T} \ln {\mathcal T} - {\mathcal R} \ln {\mathcal R}
}{{\mathcal T}}
=
\langle \! \langle \eta \rangle \! \rangle
\, . 
\label{maccopeneind}
\end{align}

\begin{figure}[ht]
\includegraphics[width=0.7 \columnwidth]{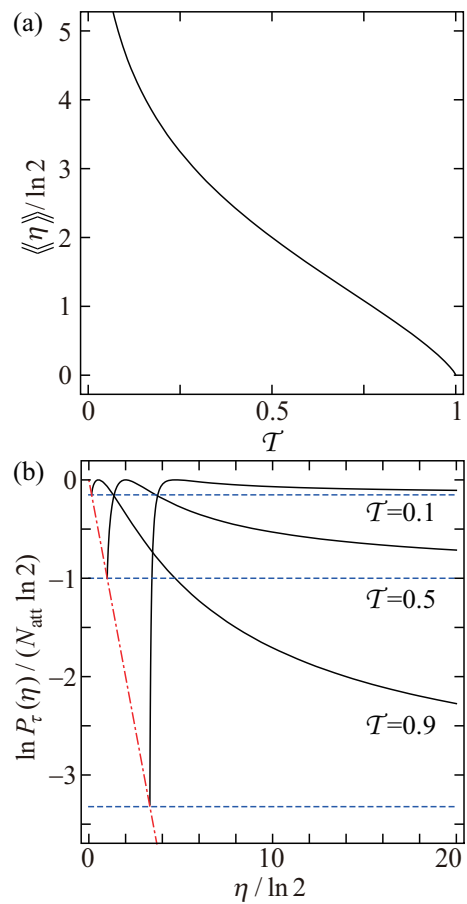}
\caption{
(a) The transmission probability dependence of the average of the COP in the limit of long measurement time $N_{\rm att} \gg 1$ (\ref{maccopeneind}). 
(b) The probability distribution of the COP for various transmission probabilities (${\mathcal T}=0.1,0.5,0.9$). 
The dot-dashed line indicates $\ln P_\tau(\eta) = - N_{\rm att} \eta$. 
The dashed lines indicate $\ln P_\tau(\eta_{\rm max}) \approx N_{\rm att} \ln {\mathcal R}$. 
}
\label{peffi}
\end{figure}

Figure \ref{peffi}(a) shows the transmission probability dependence of the average COP (\ref{maccopeneind}). 
It vanishes at ${\mathcal T}=1$ and diverges at ${\mathcal T}=0$. 
Figure \ref{peffi} (b) shows the probability distribution of COP for various transmission probabilities. 
For large $\eta$, the probability distribution approaches $\ln P(\eta_{\rm max}) \approx N_{\rm att} \ln {\mathcal R}$. 
When $\eta$ approaches the lower bound $\eta_{\rm min}$, the probability distribution approaches $\ln P(\eta_{\rm min}) \approx N_{\rm att} \, \ln {\mathcal T}=-N_{\rm att} \, \eta_{\rm min}$. 
As the transmission probability decreases, the peak position corresponding to the average value shifts rightwards, as we can expect from panel (a). 
At the same time, the tail of the probability distribution grows. 
This implies a trade-off between the amount of information content carried by a single electron and its uncertainty.

\begin{figure}[ht]
\includegraphics[width=0.7 \columnwidth]{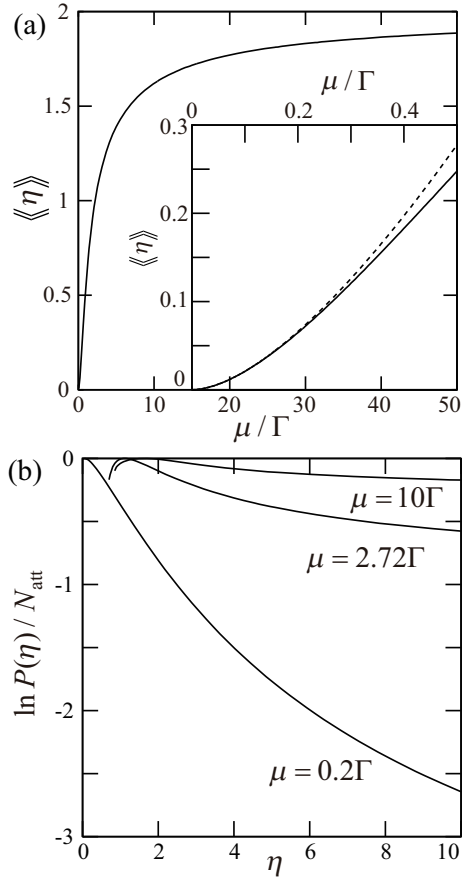}
\caption{
(a) Average of the COP as a function of the bias voltage for symmetric case; $\Gamma_L=\Gamma_R$ and $\mu_R-\epsilon_D=\epsilon_D-\mu_L=\mu/2$. 
The inset shows a magnification close to the zero bias voltage. 
The dashed line indicates the approximate expression~(\ref{copsmall}). 
At a large bias voltage, the COP becomes saturated at $\langle \! \langle \eta \rangle \! \rangle=2$. 
(b) Probability distributions of the COP for various bias voltages ($v=\mu/\Gamma=0.2,2.72,10$). 
}
\label{peta}
\end{figure}

Equation (\ref{pdfVerley}) is applicable to the energy-dependent transmission case, see Sec.~\ref{edtp}.  
Figure \ref{peta} (a) is the average COP as a function of bias voltage. 
At small bias voltages $v = \mu/\Gamma \ll 1$, by using Eqs.~(\ref{c1nsmall}) and (\ref{c1ismall}), the average COP at the steady state is approximately calculated as [dashed line in the inset of panel (a)], 
\begin{align}
\langle \! \langle \eta \rangle \! \rangle
\approx
v^2 (5/3-2 \ln v)/(3-v^2)
\, , 
\label{copsmall}
\end{align}
which increases as $\sim -v^2 \ln v$. 
As the bias voltage increases, the average COP increases and becomes saturated at, 
\begin{align}
\langle \! \langle \eta \rangle \! \rangle=2
\, , 
\label{copslarge}
\end{align}
which can be derived by using Eq.~(\ref{aveinc}). 
This $\Gamma$-independent value is the information content carried by a single electron in the incoherent sequential tunneling regime. 
Figure \ref{peta} (b) is the probability distribution of COP for various bias voltages. 
It exhibits a tendency similar to that observed in Fig.~\ref{peffi} (b), i.e., the trade-off between the amount of information content carried by a single electron and its uncertainty. 

\section{Summary}
\label{summary}

In the present paper, we extended the multi-contour Keldysh Green function technique and calculated the R\'enyi entropy for a positive integer order $M$ subjected to the particle number constraint. 
By extending $M$ to a complex number, we relate the R\'enyi entropy to the information generating function, which is the Fourier transform of the joint probability distribution of self-information and particle number. 

We applied our framework to the spinless single-resonant level model. 
For the energy-independent transmission probability, we calculated the time dependence of the accessible entanglement entropy, i.e., the conditional entropy. 
We found the nonlinear time dependence of the accessible entanglement entropy, which is attributable to the time dependence of the classical Shannon entropy of the probability distribution of the particle number. 
Although the averages of the accessible entanglement entropy and the full entanglement entropy coincide at the steady state, their fluctuations behave differently. 

We analyzed the joint probability distribution. 
For energy-independent transmission, there exists a perfect linear correlation between the self-information and the particle number; for ${\mathcal R} > {\mathcal T}$ (${\mathcal R} < {\mathcal T}$), they are positively (negatively) and perfectly correlated. 
This implies that one can determine the self-information and the entanglement entropy by counting the number of electrons. 
The energy dependence of the transmission probability spoils the perfect linear correlation. 
For a symmetric condition, when the bias voltage is smaller than the coupling strength, $\mu \ll \Gamma$, the coherent resonant tunneling process is dominant and the full entanglement entropy increases as $-(\mu/\Gamma)^3 \ln (\mu/\Gamma)$. 
For a large bias voltage, $\mu \gg \Gamma$, when the incoherent sequential tunneling process is dominant, the full entanglement entropy becomes $\tau \Gamma/2$. 
A negative (positive) linear correlation between the self-information and the number of particles is observed for $\mu \ll \Gamma$ ($\mu \gg \Gamma$). 

We also considered an analogy to the Peltier effect and analyzed the efficiency of information transmission. 
We calculated the probability distribution of the COP, 
which measures the information content carried by a single electron. 
Our results demonstrated that when the average of the COP increases, the fluctuation also increases.
This implies a trade-off between the amount of information content carried by a single particle and its uncertainty.

In the present paper, we have limited our discussion to the case when the reduced density matrix commutes with the local particle number operator of the subsystem. 
Then the accessible entanglement entropy is trivially the full entanglement entropy subtracted by the Shannon entropy. 
It would be interesting to apply our method to quantum circuits including superconducting leads, in which the local particle number superselection rule would be nontrivial.

We thank Dmitry Golubev, Hiroaki Okada, and Kazutaka Takahashi for valuable discussions. 
This work was supported by JSPS KAKENHI Grants No. 26400390 and No. JP26220711.

\appendix

\section{Proof of Eq.~(\ref{pnsc})}
\label{sec:prolocssr}

Superselection rules are due to some underlying symmetry of the system. 
Here we consider the following conditions. 
(i) The {\it total} particle number is conserved, 
\begin{align}
[ \hat{H}, \hat{N}_A+\hat{N}_B ] =0 \, . 
\label{a1}
\end{align}
(ii) The initial state (\ref{rhoini}) is diagonal in the particle number sector, 
\begin{align}
[\hat{\rho}_{\rm eq},\hat{N}_A+\hat{N}_B]=0 \, . 
\label{a2}
\end{align}
These conditions are fulfilled for nonsuperconducting leads. 
The lhs of Eq.~(\ref{pnsc}) is calculated as, 
\begin{align}
[ \hat{\rho}_A(\tau), \hat{N}_A ] 
=&
{\rm Tr}_B 
\left(
 [ 
e^{-i \hat{H} \tau} \hat{\rho}_{{\rm eq}} e^{i \hat{H} \tau}
, \hat{N}_A + \hat{N}_B ]
\right)
\nonumber \\
&-
{\rm Tr}_B 
\left(
 [ 
e^{-i \hat{H} \tau} \hat{\rho}_{{\rm eq}} e^{i \hat{H} \tau}
, \hat{N}_B ]
\right)
 \, . 
\label{proof1}
\end{align}
By exploiting Eqs.~(\ref{a1}) and (\ref{a2}), one can check that the first line of the rhs of Eq.~(\ref{proof1}) is zero. 
The second line of the rhs of Eq.~(\ref{proof1}) is also zero because of the cyclic property of the partial trace over subsystem $B$; 
\begin{align}
{\rm Tr}_B \left( \hat{N}_B e^{-i \hat{H} \tau} \hat{\rho}_{{\rm eq}} e^{i \hat{H} \tau} \right)
=
{\rm Tr}_B \left(e^{-i \hat{H} \tau} \hat{\rho}_{{\rm eq}} e^{i \hat{H} \tau} \hat{N}_B \right) \, . 
\end{align}
These discussions prove Eq.~(\ref{pnsc}). 

It would be straightforward to generalize the above proof to any conserved quantity. 
We point out that the proof generalizes a known property of the real-time diagrammatic technique~\cite{SchoellerPRB1994,KoenigPRB1996,YU2005}; if the initial reduced density matrix is diagonal in the space of a conserved quantity, e.g., the spin space, the reduced density matrix will be diagonal at all times in this space~\cite{KoenigPRB1996}. 
Obviously, the proof is applicable to both fermions and bosons. 
However, the proof is not valid when the initial state is a superposition of different particle number states, such as the BCS state~\cite{Beenakker}, and the boson coherent state~\cite{AharonovPR1967,Aharonov}.

\section{R\'enyi entropy for decoupled subsystems}
\label{sec:renentdec}

Here we analyze the modified R\'enyi entropy for the dot (\ref{rendotchi}) in the limit of zero temperature. 
The modified R\'enyi entropies of leads can be treated in the same way. 
By substituting Eq.~(\ref{rendotchi}) to Eq.~(\ref{renss}), and by exploiting Eq.~(\ref{igfbi}), the R\'enyi entropy for the dot is calculated as, 
$S_{D \, M}(N_A) = \delta_{N_A,0} {f_D^-}^M  + \delta_{N_A,1} {f_D^+}^M$. 
In the following two procedures, let us calculate the sizes of available states in the Fock subspace containing $N_A$ particles (\ref{jarequ}) in the limit of zero temperature. 
\begin{enumerate}
\item
First, we fix a finite temperature and extend $M \to 1 - i \xi$. 
The joint probability distribution becomes, 
\begin{align}
P_D(I_A^\prime,N_A)
=&
\int \frac{d \xi}{2 \pi}
e^{-i \xi I_A^\prime}
S_{D \, 1-i \xi}(N_A)
\nonumber \\
=&
f_D^- 
\, \delta_{N_A,0} \, \delta( I_A^\prime + \ln f_D^-)  
\nonumber \\
&+ 
f_D^+
\, \delta_{N_A,1} \, \delta( I_A^\prime + \ln f_D^+)
\, . 
\label{pd1}
\end{align}
Thus the `Jarzynski equality' (\ref{jarequ}) indicates that the size of the available states for fixed $N_A$ is,
$ S_{D \, 0}(N_A) = \delta_{N_A,0} + \delta_{N_A,1}=1$, 
which is temperature independent. 

\item
We first take the zero temperature limit for a positive integer $M$, and then extend $M$ to a complex number. 
The joint probability distribution is, 
\begin{align}
P_D(I_A^\prime,N_A)
=&
\delta(I_A^\prime)
\, 
[\delta_{N_A, 0}
\theta (\epsilon_D-\mu_D)
\nonumber \\
&+
\delta_{N_A, 1}
\theta (\mu_D - \epsilon_D)
]
\, . 
\label{pd2}
\end{align}
From the `Jarzynski equality' (\ref{jarequ}), we obtain the size of the available states for fixed $N_A$ as,
$S_{D \, 0}(N_A) = \delta_{N_A, 0} \theta (\epsilon_D-\mu_D) + \delta_{N_A, 1} \theta (\mu_D - \epsilon_D)$. 
\end{enumerate}
The sizes of available states in the Fock subspace for fixed $N_A$ obtained in the two procedures are different; the former is greater than or equal to the latter. 
The difference is attributable to a delta peak of Eq.~(\ref{pd1}) at $I_A^\prime \approx \beta |\epsilon_D - \mu_D| \to \infty$ with exponentially small weight $\sim e^{-\beta |\epsilon_D - \mu_D|}$, which remains to contribute even in the limit of zero temperature.

\begin{widetext}

\section{Derivations of the multi-contour Keldysh Green function}
\label{sec:mcgf}

We calculate the components of the matrix multi-contour Keldysh Green function (\ref{matmcgf}). 
Here, we present detailed calculations of a particular component, $m > m'$, $s=-$ and $s=+$. 
Other components can be calculated in the same manner. 
Noting that the contour ordering operator $T_C$ also acts on the replicated equilibrium density matrices $\rho_{{\rm eq},m}$ ($m=1,\cdots, M$), we obtain 
\begin{align}
g_{Lk}^{ \{ \chi_j \}, m-,m'+}(t,t')
=&
-i 
{\rm Tr}
\left[
\hat{T}_C
\hat{a}_{Lk}(t_{ms})_I
\hat{a}_{Lk}^\dagger(t_{m's'}')_I
e^{i \sum_{j=1}^M \chi_j \hat{N}_L(\tau_{j+})_I}
\hat{\rho}_{{\rm eq},M}
\cdots
\hat{\rho}_{{\rm eq},1}
\right]
/s_{L,M}
\\
=&
-i 
\, 
\frac{
{\rm Tr}
\left[
e^{i \chi_M \hat{N}_L} \hat{\rho}_{{\rm eq},M}
\cdots 
\hat{\rho}_{{\rm eq},m}
\hat{a}_{L k}(t_{m})_I
e^{i \chi_{m-1} \hat{N}_L}
\cdots 
e^{i \chi_{m'} \hat{N}_L}
\hat{a}_{Lk}^\dagger(t_{m'}')_I
\hat{\rho}_{{\rm eq},m'}
\cdots 
e^{i \chi_{1} \hat{N}_L}
\hat{\rho}_{{\rm eq},1}
\right]
}
{s_{L,M}}
\\
=&
-i 
\, 
\frac{
{\rm Tr}_L
\left[
e^{i (\sum_{j=m}^M+\sum_{j=1}^{m'-1}) \chi_j \hat{N}_L}
{ \hat{\rho}_{L, {\rm eq}} }^{M-m+m'+1}
\hat{a}_{L k}(t_{m})_I
e^{i \sum_{j=m'}^{m-1} \chi_j \hat{N}_L}
{ \hat{\rho}_{L, {\rm eq}} }^{m-m'-1}
\hat{a}_{L k}^\dagger(t_{m'}')_I
\right]
}{
{\rm Tr}_L
\left[
e^{i \bar{\chi} \hat{N}_L}
{\hat{\rho}_{L, {\rm eq}}}^M
\right]
} 
\, ,
\end{align}
In the following, we will omit the subscripts $k$ and $L$. 
By using the following relation, 
\begin{align}
\frac{
{\rm Tr}
\left[
e^{i (\bar{\chi}-\Delta \chi) \hat{N}}
{\rho_{ {\rm eq}}}^{M-n}
\hat{a}
e^{i \Delta \chi \hat{N}}
{\hat{\rho}_{ {\rm eq}}}^n
\hat{a}^\dagger 
\right]
}{
{\rm Tr}
\left[
e^{i \bar{\chi} \hat{N}}
{\hat{\rho}_{{\rm eq}}}^M
\right]
}
=
\frac{
e^{i \Delta \chi - n \beta (\epsilon-\mu) }
}{
1+e^{i \bar{\chi} - M \beta (\epsilon-\mu)}}
=
e^{i (\Delta \chi - n \bar{\chi}/M) }
f_{n}^{\bar{\chi}}(\epsilon)
\, , 
\end{align}
we obtain,  
\begin{align}
g^{\{ \chi_j \}, m-,m'+}(t,t')
=
-i f_{m-m'-1}^{\bar{\chi}}(\epsilon) e^{-i \epsilon (t-t')
+
i \sum_{j=m'}^{m-1} \delta \chi_j
+
i \bar{\chi}/M
}
\, . 
\end{align}
Three other components for $m > m'$, are calculated in the following: 
\begin{align}
g^{\{ \chi_j \}, m+,m'-}(t,t')
=&
-i 
\, 
\frac{
{\rm Tr}
\left[
e^{i (\sum_{j=m+1}^M + \sum_{j=1}^{m'-1}) \chi_j \hat{N}}
{\hat{\rho}_{\rm eq}}^{M-m+m'-1}
\hat{a}(t_{m})_I
e^{i \sum_{j=m'}^{m-1} \chi_j \hat{N}}
{\hat{\rho}_{\rm eq}}^{m-m'+1}
\hat{a}^\dagger(t_{m'}')_I
\right]
}{
{\rm Tr} \left[ e^{i \bar{\chi} \hat{N}} {\hat{\rho}_{{\rm eq}}}^M \right]
}
\nonumber \\
&=
-i f_{m-m'+1}^{\bar{\chi}}(\epsilon) e^{-i \epsilon (t-t')
+
i \sum_{j=m'}^{m-1} \delta \chi_j
-
i \bar{\chi}/M
}
\, ,
\\
g^{\{ \chi_j \} , m \pm, m' \pm}(t,t')
=&
-i e^{-i \epsilon (t-t')}
\, 
\frac{
{\rm Tr}
\left[
e^{i (\sum_{j=m}^M+\sum_{j=1}^{m'-1}) \chi_j \hat{N}}
{\hat{\rho}_{\rm eq}}^{M-m+m'}
\hat{a}
e^{i \sum_{j=m'}^{m-1} \chi_j \hat{N}}
{\hat{\rho}_{\rm eq}}^{m-m'}
\hat{a}^\dagger
\right]
}{
{\rm Tr} \left[ e^{i \bar{\chi} \hat{N}} {\hat{\rho}_{{\rm eq}}}^M \right]
}
\nonumber \\
&=
-i f^{\bar{\chi}}_{m-m'}(\epsilon) e^{-i \epsilon (t-t')
+
i \sum_{j=m'}^{m-1} \delta \chi_j
}
\, . 
\end{align}
Four components for $m<m'$ are, 
\begin{subequations}
\begin{align}
g^{\{ \chi_j \} , m-,m'+}(t,t')
=&
i e^{-i \epsilon (t-t')}
\, 
{\rm Tr}
\left[
e^{i \sum_{j=m}^{m'-1} \chi_j  \hat{N}}
{\hat{\rho}_{\rm eq}}^{m'-m+1}
\hat{a}
e^{i (\bar{\chi} - \sum_{j=m}^{m'-1}) \chi_j \hat{N}}
{\hat{\rho}_{\rm eq}}^{M-m'+m-1}
\hat{a}^\dagger
\right]
/{\rm Tr} \left[ e^{i \bar{\chi} \hat{N}} {\hat{\rho}_{{\rm eq}}}^M \right]
\nonumber \\
=&
i f^{\bar{\chi}}_{M-m'+m-1}(\epsilon) e^{-i \epsilon (t-t')
-
i \sum_{j=m}^{m'-1} \delta \chi_j
+
i \bar{\chi}/M
}
\, ,
\\
g^{\{ \chi_j \} , m+,m'-}(t,t')
=&
i e^{-i \epsilon (t-t')}
\, 
{\rm Tr}
\left[
e^{i \sum_{j=m}^{m'-1} \chi_j \hat{N}}
{\hat{\rho}_{\rm eq}}^{m'-m-1}
\hat{a}
e^{i (\bar{\chi}- \sum_{j=m}^{m'-1} \chi_j) \hat{N}}
{\hat{\rho}_{\rm eq}}^{M-m'+m+1}
\hat{a}^\dagger
\right]
/{\rm Tr} \left[ e^{i \bar{\chi} \hat{N}} {\hat{\rho}_{{\rm eq}}}^M \right]
\nonumber \\
=&
i f^{\bar{\chi}}_{L,M-m'+m+1}(\epsilon) e^{-i \epsilon (t-t')
-
i \sum_{j=m}^{m'-1} \delta \chi_j
-
i \bar{\chi}/M
}
\, ,
\\
g^{\{ \chi_j \} , m \pm ,m' \pm}(t,t')
=&
i e^{-i \epsilon (t-t')}
\, 
{\rm Tr}
\left[
e^{i \sum_{j=m}^{m'-1} \chi_j \hat{N}}
{ \hat{\rho}_{\rm eq} }^{m'-m}
\hat{a}
e^{i (\bar{\chi} - \sum_{j=m}^{m'-1} \chi_j) \hat{N}}
{ \hat{\rho}_{\rm eq} }^{M-m'+m}
\hat{a}^\dagger
\right]
/{\rm Tr} \left[ e^{i \bar{\chi} \hat{N}} {\hat{\rho}_{{\rm eq}}}^M \right]
\nonumber \\
=&
i 
f^{\bar{\chi}}_{M-m'+m}(\epsilon) 
e^{-i \epsilon (t-t') -i \sum_{j=m}^{m'-1} \delta \chi_j }
\, . 
\end{align}
\end{subequations}
Four components defined on the same replica $m=m'$ are, 
\begin{subequations}
\begin{align}
g^{\{ \chi_j \} , m+,m-}(t,t')
=&
i e^{-i \epsilon (t-t')}
\, 
{\rm Tr}
\left[
e^{i {\bar{\chi}} \hat{N}}
{\hat{\rho}_{\rm eq}}^{M-1}
\hat{a}
{\hat{\rho}_{\rm eq}}
\hat{a}^\dagger
\right]
/{\rm Tr} \left[ e^{i {\bar{\chi}} \hat{N}} {\hat{\rho}_{{\rm eq}}}^M \right]
=
-
i f^{\bar{\chi}}_{1}(\epsilon) e^{-i \epsilon (t-t')-i {\bar{\chi}}/M}
\, ,
\\
g^{\{ \chi_j \} , m-,m+}(t,t')
=&
i e^{-i \epsilon (t-t')}
\, 
{\rm Tr}
\left[
e^{i {\bar{\chi}} \hat{N}}
{\hat{\rho}_{\rm eq}}^{M-1}
\hat{a}^\dagger
{\hat{\rho}_{\rm eq}}
\hat{a}
\right]
/{\rm Tr} \left[ e^{i {\bar{\chi}} \hat{N}} {\hat{\rho}_{{\rm eq}}}^M \right]
=
i f^{\bar{\chi}}_{M-1}(\epsilon) e^{-i \epsilon (t-t')+i {\bar{\chi}}/M}
\, ,
\\
g^{\{ \chi_j \} , m + ,m' +}(t,t')
=&
-i e^{-i \epsilon (t-t')}
\,
\left \{
\theta(t-t') 
{\rm Tr}
\left[
e^{i {\bar{\chi}} \hat{N}}
{\hat{\rho}_{\rm eq}}^{M} 
\hat{a} 
\hat{a}^\dagger
\right]
-
\theta(t'-t) 
{\rm Tr}
\left[
e^{i {\bar{\chi}} \hat{N}}
{\hat{\rho}_{\rm eq}}^{M} 
\hat{a}^\dagger 
\hat{a}
\right]
\right \}
/{\rm Tr} \left[ e^{i {\bar{\chi}} \hat{N}} 
{\hat{\rho}_{{\rm eq}}}^M \right]
\nonumber \\
=&
-i 
e^{-i \epsilon (t-t')}
\left[
\theta(t-t')  f^{\bar{\chi}}_{0}(\epsilon)
-
\theta(t'-t)  f^{\bar{\chi}}_{M}(\epsilon)
\right]
\, , 
\\
g^{\{ \chi_j \} , m - ,m' -}(t,t')
=&
-i 
e^{-i \epsilon (t-t')}
\left[
\theta(t'-t)  f^{\bar{\chi}}_{0}(\epsilon)
-
\theta(t-t')  f^{\bar{\chi}}_{M}(\epsilon)
\right]
\, . 
\end{align}
\end{subequations}

\section{Discrete Fourier transform}
\label{sec:dft}

Here we present detailed calculations on the discrete Fourier transform~(\ref{dft}) for $\delta \chi_j=0$. 
In the following, we put $x_\ell=e^{i \pi (2 \ell+1)/M}=- e^{-i \lambda_\ell}$
and use the same notations in Appendix~\ref{sec:mcgf}. 
The discrete Fourier transform of the `greater' component is calculated as follows:  
\begin{align}
\sum_{m=1}^M
g^{\{ \chi/M \} , m-,m'+}
e^{i \pi \frac{2 \ell+1}{M} (m-m')}
=&
-i e^{-i \epsilon(t-t')+i \chi/M}
\nonumber \\
& \times
\left(
\sum_{m=m'+1}^M
f_{m-m'-1}^\chi(\epsilon)
x_\ell^{m-m'}
-
f_{M-1}^\chi(\epsilon)
-
\sum_{m=1}^{m'-1}
f_{M+m-m'+1}^\chi(\epsilon)
x_\ell^{m-m'}
\right)
\nonumber \\
=&
-i e^{-i \epsilon(t-t')+i \chi/M}
\sum_{j=1}^{M}
f_{L,j-1}^\chi(\epsilon)
x_\ell^j
=
i e^{-i \epsilon(t-t')-i(\lambda_\ell-\chi/M)}
f^-_{L, \lambda_\ell-\chi/M}(\epsilon)
\nonumber \\
=&
g^{\lambda_\ell-\chi/M,-+}(t,t')
\, . 
\end{align}
%
The discrete Fourier transform of the `lesser' component is, 
\begin{align}
\sum_{m=1}^M
g^{\{ \chi/M \} , m+,m'-}
e^{i \pi \frac{2 \ell+1}{M} (m-m')}
=&
-i e^{-i \epsilon(t-t')-i \chi/M}
\left(
\sum_{m=m'+1}^M
f_{m-m'+1}^\chi(\epsilon)
x_\ell^{m-m'}
+
f_{1}^\chi(\epsilon)
-
\sum_{m=1}^{m'-1}
f_{M+m-m'+1}^\chi(\epsilon)
x_\ell^{m-m'}
\right)
\nonumber \\
=&
-i e^{-i \epsilon(t-t')-i \chi/M}
\sum_{j=0}^{M-1}
f_{j+1}^\chi(\epsilon)
x_\ell^j
=
-i e^{-i \epsilon(t-t')+i(\lambda_\ell-\chi/M)}
f^+_{\lambda_\ell-\chi/M}(\epsilon)
\nonumber \\
=&
g^{\lambda_\ell-\chi/M,+-}(t,t')
\, . 
\end{align}
The discrete Fourier transform of the causal (anti-causal) component is, 
\begin{align}
\sum_{m=1}^M
g^{\{ \chi/M \} , m \pm,m' \pm}
e^{i \pi \frac{2 \ell+1}{M} (m-m')}
=&
-i e^{-i \epsilon(t-t')}
\biggl(
\sum_{m=m'+1}^M
f_{m-m'}^\chi(\epsilon)
x_\ell^{m-m'}
+
f_{0}^\chi(\epsilon)
\theta (\pm(t-t'))
\nonumber \\
&
-
f_{M}^\chi(\epsilon)
\theta (\pm(t'-t))
-
\sum_{m=1}^{m'-1}
f_{M+m-m'}^\chi(\epsilon)
x_\ell^{m-m'}
\biggl)
\nonumber \\
=&
-i e^{-i \epsilon(t-t')}
\left(
\theta (\pm(t-t'))
\sum_{j=0}^{M-1}
f_{j}^\chi(\epsilon) x_\ell^j
+
\theta (\pm(t'-t))
\sum_{j=1}^{M}
f_{j}^\chi(\epsilon) x_\ell^j
\right)
\nonumber \\
=&
-i e^{-i \epsilon(t-t')}
\left[
f_{\lambda_\ell-\chi/M}^-(\epsilon)
\theta( \pm (t-t') )
-
f_{\lambda_\ell-\chi/M}^+(\epsilon)
\theta( \pm (t'-t) )
\right]
\nonumber \\
=&
g^{\lambda_\ell-\chi/M,\pm \pm}(t,t')
\, . 
\end{align}

\section{Boundary points}
\label{bps}

For $v<1$, the asymptotic form of the R\'enyi entropy (\ref{igfrlm}) for $i \xi \to \pm \infty$ ($M \to \mp \infty$) is
\begin{subequations}
\begin{align}
\ln S_M(\lambda)
\approx & 
M \frac{\tau}{2 \pi}
\int_{\mu_L}^{\mu_R}
\ln
{\mathcal R}(\omega)
+ i \chi N_{A,0}
=
-
N_{\rm att}
M 
(F_1(0)-\ln (v/e)^2)
+ i \chi N_{A,0}
\, , 
\;\;\;\;
i \xi \to \infty \; (M \to -\infty)
\, , 
\\
\ln S_M(\lambda)
\approx&
M \frac{\tau}{2 \pi}
\int_{\mu_L}^{\mu_R} d \omega
\ln
{\mathcal T}(\omega)
e^{i \chi/M}
+ i \chi N_{A,0}
=
N_{\rm att}
(i \chi -M F_1(0))
+ i \chi N_{A,0}
\, , 
\;\;\;\;
i \xi \to - \infty \; (M \to \infty)
\, . 
\end{align}
\end{subequations}
Then, by exploiting the Legendre duality~\cite{Touchette}, boundary points of support with maximum and maximum $I_A$, ${\mathbf Q}$ and ${\mathbf Q}'$, for $v<1$, are obtained as, 
\begin{subequations}
\begin{align}
{\mathbf Q}
=&
\lim_{i \xi \to + \infty}
(
\partial_{i \xi} \ln S_{1-i \xi}(\chi)
,
\partial_{i \chi} \ln S_{1-i \xi}(\chi)
)
=
(N_{\rm att} (F_1(0)-\ln(v/e)^2),N_{A,0})
\, , 
\\
{\mathbf Q}'
=&
\lim_{i \xi \to - \infty}
(
\partial_{i \xi} \ln S_{1-i \xi}(\chi)
,
\partial_{i \chi} \ln S_{1-i \xi}(\chi)
)
=
(N_{\rm att} F_1(0),N_{\rm att}+N_{A,0})
\, . 
\end{align}
\end{subequations}

For $v>1$, we have to pay attention to the condition that ${\mathcal T}(\omega)>{\mathcal R}(\omega)$ for $|\omega-\epsilon_D| < \Gamma/2$ and ${\mathcal T}(\omega)<{\mathcal R}(\omega)$ for $|\omega-\epsilon_D| > \Gamma/2$, see Eq.~(\ref{sc}). 
The asymptotic form of the R\'enyi entropy (\ref{igfrlm}) is, 
\begin{subequations}
\begin{align}
\ln S_M(\chi)
\approx&
M \frac{\tau}{2 \pi}
\int_{\epsilon_D - \Gamma/2}^{\epsilon_D + \Gamma/2} d \omega
\ln
{\mathcal R}(\omega)
+
M \frac{\tau}{2 \pi}
\left(
\int_{\epsilon_D + \Gamma/2}^{\mu_R} d \omega
+
\int_{\mu_L}^{\epsilon_D - \Gamma/2} d \omega
\right)
\ln
{\mathcal T}(\omega)
e^{i \chi/M}
+ i \chi N_{A,0}
\nonumber \\
=&
i \chi N_{\rm att} (1-1/v)
-
M N_{\rm att} (F_1(0) + 2/v)
+ i \chi N_{A,0}
\, , 
\;\;\;\;
i \xi \to \infty \; (M \to - \infty)
\, , 
\\
\ln S_M(\chi)
\approx&
\frac{\tau}{2 \pi}
\int_{\epsilon_D - \Gamma/2}^{\epsilon_D + \Gamma/2} d \omega
\ln
{\mathcal T}(\omega)^M
e^{i \chi}
+
M
\frac{\tau}{2 \pi}
\left(
\int_{\epsilon_D + \Gamma/2}^{\mu_R} d \omega
+
\int_{\mu_L}^{\epsilon_D - \Gamma/2} d \omega
\right)
\ln
{\mathcal R}(\omega)
+ i \chi N_{A,0}
\nonumber \\
=&
i \chi  
N_{\rm att}
/v
-
N_{\rm att}
M
[F_1(0)-\ln (v/e)^2-2/v]
+ i \chi N_{A,0}
\, , 
\;\;\;\;
i \xi \to - \infty \; (M \to \infty)
\, . 
\end{align}
\end{subequations}
Then the Legendre duality~\cite{Touchette} implies that boundary points of support with maximum and maximum $I_A^\prime$, ${\mathbf Q}$ and ${\mathbf Q}'$, are 
\begin{subequations}
\begin{align}
{\mathbf Q}
=&
\lim_{i \xi \to + \infty}
(
\partial_{i \xi} \ln S_{1-i \xi}(\chi)
,
\partial_{i \chi} \ln S_{1-i \xi}(\chi)
)
=
(N_{\rm att} [F_1(0)+2/v],N_{\rm att} (1-1/v)+N_{A,0})
\, , 
\\
{\mathbf Q}'
=&
\lim_{i \xi \to - \infty}
(
\partial_{i \xi} \ln S_{1-i \xi}(\chi)
,
\partial_{i \chi} \ln S_{1-i \xi}(\chi)
)
=
(N_{\rm att} [F_1(0)-\ln (v/e)^2-2/v],N_{\rm att}/v+N_{A,0})
\, . 
\end{align}
\end{subequations}

\end{widetext}


\begin{thebibliography}{99}
\bibitem{note1}
The standard definition of the R\'enyi entropy of order $M$ is 
$(\ln {\rm Tr}_A
\left[
{\rho_A}^M
\right]
)/(1-M)$ for $M \geq 0$ and $M \neq 1$ , see, e.g., Ref.~\onlinecite{Cover}. 
In the present paper, we adopt the terminology in Ref.~\onlinecite{Nazarov}.
\end{thebibliography}
\end{document}